\documentclass[a4paper, 11pt]{article}
\usepackage[utf8]{inputenc}
\usepackage[]{graphicx}
\usepackage[dvipsnames,table]{xcolor}
\usepackage{alltt}
\PassOptionsToPackage{hyphens}{url}\usepackage{hyperref}
\usepackage{amsmath, amssymb}
\usepackage{doi} 
\usepackage[round, authoryear]{natbib} 
\usepackage{multirow}
\usepackage{times}
\usepackage{comment}
\usepackage{caption}
\usepackage{subcaption}
\usepackage{enumitem} 
\usepackage{tabularx}
\usepackage{bbm}
\usepackage{algorithm}
\usepackage{algpseudocode}
\usepackage{microtype}
\usepackage{booktabs} 
\usepackage[title]{appendix} 
\usepackage{nameref} 
\usepackage[dvipsnames,table]{xcolor}
\usepackage[onehalfspacing]{setspace} 
\usepackage[labelfont=bf,font=small]{caption} 
\usepackage{newtxtext,newtxmath}
\usepackage{sectsty} 
\allsectionsfont{\sffamily} 
\usepackage[labelfont={bf,sf},font=small]{caption} 
\usepackage{orcidlink} 
\usepackage{fancyhdr}
\usepackage{geometry}
\usepackage{siunitx}%
\usepackage{adjustbox}
\usepackage{bm}
\usepackage{etoolbox}
\usepackage{placeins}

\newrobustcmd\B{\DeclareFontSeriesDefault[rm]{bf}{b}\bfseries}

\makeatletter
\def\maxwidth{ %
  \ifdim\Gin@nat@width>\linewidth
    \linewidth
  \else
    \Gin@nat@width
  \fi
}
\makeatother

\usepackage{framed}
\makeatletter
 {\par\unskip\endMakeFramed%
 \at@end@of@kframe}
\makeatother

\definecolor{unipd_red}{HTML}{b40908}

\newcommand\mail{roberto.macridemartino@deams.units.it}
\title{\vspace{-2em}
  \textbf{\textbf{Bayesian weighted discrete-time dynamic models for association football prediction}}
}
\author{
     \textbf{Roberto Macrì-Demartino}    \orcidlink{0000-0002-5296-6566} \thanks{Corresponding author e-mail: \href{mailto:\mail}{\texttt{\mail}}} , \textbf{Leonardo Egidi} \orcidlink{0000-0003-3211-905X}, and \textbf{Nicola Torelli} \\
  \small Department of Economics, Business, Mathematics, and Statistics ``Bruno de Finetti", University of Trieste,\\
  \small Via A. Valerio 4/1, Trieste, 34127, Italy\vspace{2em} \\
  {\color{unipd_red}\small {THIS IS A PREPRINT WHICH HAS NOT YET BEEN PEER REVIEWED}}
}
\date{}

\IfFileExists{upquote.sty}{\usepackage{upquote}}{}

\usepackage{hyperref}
\hypersetup{
  bookmarksopen=true,
  breaklinks=true,
  colorlinks=true,
  linkcolor=blue,
  anchorcolor=black,
  citecolor=blue,
  urlcolor=black,
}

\newcommand{\indep}{\perp \!\!\! \perp}

\begin{document}

\maketitle

\begin{abstract}
In recent years, great emphasis has been placed on the prediction of association football. Due to this,
several studies have proposed different types of statistical models to predict the outcome of a football match. However, most existing approaches usually assume that the offensive and defensive abilities of teams remain static over time. We introduce a Bayesian dynamic approach for football goal-based models that uses period-specific commensurate priors to flexibly weight the evolution of attacking and defensive abilities. Our approach assigns separate, time-varying precisions for each ability and period, controlled via spike-and-slab hyperpriors. This adaptive shrinkage borrows information about teams' strength when past and current performance aligns and allows rapid adjustments when teams experience substantial changes (e.g., transfer windows or coaching changes). We integrate this framework into six standard goal-based models evaluating predictive performance using data from the last five seasons of the German Bundesliga, English Premier League, and Spanish La Liga. Compared with the other discrete‑time dynamic models, our adaptive approach yields better predictive performance. The proposed
methodology has also been implemented in the free and open source R package \texttt{footBayes}.
  \\[1ex]
\end{abstract}

\maketitle

\section{Introduction}\label{sec1}
Quantitative analysis of association football (soccer), hereafter referred to as football, and more specifically, matches' prediction is a rapidly evolving discipline, increasingly valued by participants, coaches, owners, and gamblers looking to gain a competitive advantage. Consequently, there is a growing demand for information that supports better decision making. 

The outcome of football matches can be predicted using two main statistical modelling frameworks. In the goal-based (or direct) approach, the actual numbers of goals scored by each team is a count variable -- most commonly modelled via Poisson or Negative-Binomial regression. The expected goal counts are functions of team attributes (e.g., offensive and defensive abilities) and, when relevant, home-field advantage. In contrast, result-based (or indirect) models predict one of three outcomes -- home win, draw, or away win -- typically through ordered probit \citep{koning2000} or logit \citep{carpita2015discovering, Carpita2019} regressions. 
Furthermore, the widespread popularity of large datasets promoted the use of machine learning (ML) tools, yielding a fundamentally different modelling approach based on a random forest \citet{breiman2001random}. The potential of using this as a new results-based model was first explored by \citet{Schauberger_Groll2018} to assess the predictive performance of different types of random forests compared to classical Poisson regression methods on data containing all matches of the FIFA World Cups 2002–2014. Along these lines, \citet{groll2019}, \citet{groll2019hybrid}, \citet{groll2021}, and \citet{groll2024modeling} further expand this framework.
It is worth noting that the result-based framework is formally nested within the goal-based one. Specifically, match results are derived from the underlying goal counts, while knowing only the three-way result provides no information about the actual goals scored, potentially misestimating team strength \citep{egidi_torelli_2021}. Therefore, we focus on the richer goal-based structure, which not only yields three-way predictions but also captures the magnitude of outcomes.

In the simplest goal-based formulation, team-specific goal counts are assumed conditionally independent given team abilities or covariates, resulting in a double Poisson model \citep[among others]{maher1982, Baio_Blangiardo2010, groll_abedieh_2013, egidi_etal2018}. To relax the strong independence assumption, several generalisations introduce score dependence. \citet{dixon_coles1997} extended the work of \citet{maher1982} by allowing (a slightly negative) correlation between scores and incorporating a dependence parameter into their model to account for it. \citet{karlis_ntzoufras2003} introduced in a frequentist framework the bivariate Poisson model, designed to account for positive goal dependencies. Furthermore, \citet{ntzoufras2011bayesian} extended it from a Bayesian perspective. 

A main assumption of previous models is the invariance of team-specific parameters, implying static offensive and defensive abilities over time.  However, it is recognised that team performance is inherently dynamic, fluctuating over years and possibly within seasons. One simple approach is the time decay weighting used by \citet{dixon_coles1997}, in which older match outcomes are downweighted so that recent games have more influence on estimated abilities. However, a more formal approach is to treat team abilities as time-varying parameters. Specifically, a continuous-time dynamic extension of the double Poisson model was introduced by \citet{Rue_2000}, while \citet{Koopman_2015} and \citet{koopman2019forecasting} integrated bivariate Poisson models in a state-space framework, allowing the abilities of the team to vary according to a state vector. Alternatively, \citet{owen_2011} proposed a Bayesian discrete-time approach based on an evolution component that describes the stochastic behaviour of time-dependent parameters. Here, the evolution component is
specified as a random walk prior distribution structure for both the attack and defence parameters.  Recent developments include \citet{egidi_etal2018}, which incorporates betting odds and other refinements, and \citet{macri2024alternative}, that evaluate predictive performance improvements when the ranking of a team is added as a covariate in dynamic models.
However, a key limitation of the approach proposed by \citet{owen_2011} is the assumption of a constant and common evolution precision for both attack and defence parameters. This constraint may limit the predictive performance of the goal-based model, ignoring the fact that a team's performance can fluctuate more during certain periods (e.g., early season transfers, midseason managerial changes). Furthermore, forcing the same evolution precision for both attack and defence neglects that these abilities can have different rates of change -- defensive abilities may adapt more slowly than offensive ones, or vice versa. 

In the present work, we try to fill this gap by proposing a Bayesian weighted discrete-time approach that incorporates time-specific commensurate priors \citep{hobbs2011hierarchical,Hobbs2012} for both attack and defence parameters. This provides a formal mechanism for letting the prior distributions at a specific time to adaptively borrow information from the previous time, but only to the extent that the data support it. By introducing flexible and dynamic evolution precision, we obtain a more accurate and adaptive modelling of team abilities over time, improving the predictive performances.

The paper is organised as follows. Section \ref{sec2} presents the Poisson and Negative Binomial goal-based models used in this study. Furthermore, Section \ref{sec3} describes the commensurate prior framework and introduces our proposed dynamic weighted approach for the offensive and defensive abilities of the teams in the goal-based models. In Section \ref{sec4}, we apply our methodology to some of the top European leagues,
namely: the German Bundesliga,  English Premier League (EPL), and Spanish La Liga.
A total of five seasons from 2020 to 2025 are used from
each league to perform the study.  Finally, Section \ref{sec5} provides concluding remarks that outline limitations, advantages, and potential future research directions.

\section{Goal-based models}\label{sec2}
This section presents the statistical goal-based models used to predict the chosen competition outcomes. Through an in-depth analysis, our aim is to provide a comprehensive overview of the methodologies applied to predict football matches, highlighting both their statistical foundations and practical implementations in sports analytics.

\subsection{Poisson-based models}\label{sec_poisson_based}
Let $(x_{i,n}, y_{j,n})$ represent the observed number of goals scored by the home and the away team in the \textit{n}-th match, with $i\neq j = 1, \dots, N_T $ and $n = 1,\dots, N$. A simple double Poisson (DP) model \citep{maher1982} assumes that the goal counts follow two conditionally independent Poisson distributions
\begin{equation}
\label{double_poisson}
\begin{aligned}
X_{i,n} \mid \lambda_{1,n} & \sim \operatorname{Poisson}\left(\lambda_{1, n}\right) \\
Y_{j,n} \mid \lambda_{2,n} & \sim \operatorname{Poisson}\left(\lambda_{2,n}\right) \\
X_{i,n}  \indep Y_{j,n} & \mid  \lambda_{1,n}, \lambda_{2,n},
\end{aligned}
\end{equation}
where the (non-negative) parameters $\lambda_{1,n}$ and $\lambda_{2,n}$ are the expected scoring rates of the home and away teams, respectively, in the \textit{n}-match. In Maher’s model, the rate at which a team is expected to score is a function of both its own offensive ability and the defensive ability of its opponent
\begin{equation}
\label{double_poisson_scoring}
    \begin{aligned}
        \log \left(\lambda_{1, n}\right) & =\beta_0 + \text{home} + \beta^{\operatorname{att}}_{h_n}+ \beta^{\operatorname{def}}_{a_n},  \\
\log \left(\lambda_{2, n}\right) & =\beta_0 + \beta^{\operatorname{att}}_{a_n} +\beta^{\operatorname{def}}_{h_n},
    \end{aligned}
\end{equation}
where the parameter $\beta_0$ is a common intercept, $\text{"home"}$ captures the well-known home-field advantage, and $\beta^{\operatorname{att}}$ and $\beta^{\operatorname{def}}$ represent the unknown attacking and defensive abilities of the home team $h_n$ and the away team $a_n$ in the \textit{n}-th match..

However, it is widely recognised that the scores of two competing football teams are positively correlated. Thus, the independence assumption in the double Poisson model \eqref{double_poisson} might be too restrictive. To account for this correlation, \citet{karlis_ntzoufras2003} introduced the bivariate Poisson (BP) model, which explicitly captures the dependence between goal counts. The joint distribution for the goals scored by the home and away teams under this model is given by the bivariate Poisson probability mass function
\begin{equation}\label{eq:bivariate_poisson}
\begin{aligned}
\mathbb{P}_{X_{i,n},Y_{j,n}}(x_{i,n},y_{j,n})
= & \exp\left\{-(\lambda_{1,n} + \lambda_{2,n} + \lambda_{3,n})\right\}
\frac{\lambda_{1,n}^{x_{i,n}}\,\lambda_{2,n}^{y_{j,n}}}{x_{i,n}!\,y_{j,n}!} \times \\
& \sum_{k=0}^{\min(x_{i,n},y_{j,n})}
\binom{x_{i,n}}{k}\,\binom{y_{j,n}}{k}\,k!\,\biggl(\frac{\lambda_{3,n}}{\lambda_{1,n}\,\lambda_{2,n}}\biggr)^k,
\end{aligned}
\end{equation}
where $\mathbb{E}(X_{i,n}) = \lambda_{1,n} + \lambda_{3,n}$
and $\mathbb{E}(Y_{j,n}) = \lambda_{2,n} + \lambda_{3,n}$. The parameter $\lambda_{3,n} = \text{cov}(X_{i,n},Y_{j,n})$ measures the covariance between the two goal counts, representing the dependence between the scores of the two teams. Furthermore, the scoring rates $\lambda_{1,n}$ and $\lambda_{2,n}$ are defined as in \eqref{double_poisson_scoring}. 
Additionally, in Equation \eqref{eq:bivariate_poisson}, we model the covariance $\lambda_{3,n}$ to not depend on other predictors
\begin{equation*}
\begin{aligned}
\log \left(\lambda_{3,n}\right) & =\eta_0.
\end{aligned}
\end{equation*}
The bivariate Poisson model generalizes the double Poisson model. Specifically, when $\lambda_{3,n} = 0$, the goal counts become independent, and the bivariate Poisson model reduces precisely to the double Poisson model described in \eqref{double_poisson}.

\subsection{Negative Binomial and Skellam alternatives}
Poisson models assume equal mean and variance, which may not hold in real-world football data -- especially in competitions where overdispersion (sample variance exceeds the sample mean) is observed in the number of
goals. To handle this, a common approach is to replace each Poisson marginal with a negative binomial (NB) distribution \citep{reep1971skill}. That is
\begin{equation*}
X_{i,n}\sim \mathrm{NB}(\lambda_{1,n},\gamma),\quad
Y_{j,n}\sim \mathrm{NB}(\lambda_{2,n},\gamma),
\end{equation*}
where $\lambda_{1,n}$ and $\lambda_{2,n}$ follow the same log-linear structure introduced in Section \ref{sec_poisson_based}, and $\gamma>0$ is the dispersion parameter.  The negative binomial model directly captures the overdispersion in the goal count of each team, with higher values of $\gamma$ indicating a greater inflation of variance compared to the Poisson models. 

Alternatively, \citet{karlis_ntzoufras2009} suggest using the Skellam distribution \citep{skellam1946frequency}, which directly models the difference in goal. Notably, the Skellam distribution captures not only the overdispersion but also the intrinsic dependence between the teams' scoring outcomes, without requiring explicit correlation modelling.
Specifically, let $X_{i,n}$ and $Y_{j,n}$ represent independent Poisson counts for goals scored by teams $T_i$ and $T_j$ in the \textit{n}-th match, respectively, with $i\neq j = 1, \dots, N_T $ and $n = 1,\dots, N$. The Skellam model (SM) is then given by the difference of the two goal counts 
\begin{align*}
    Z_n = X_{i,n}-Y_{j,n}.
\end{align*}
The corresponding probability mass function is given by
\begin{equation}
\label{skellam}
\begin{aligned}
\mathbb{P}_{Z_n}(z_n) =  \exp\left\{-(\lambda_{1,n} + \lambda_{2,n})\right\} \left( \frac{\lambda_{1,n}}{\lambda_{2,n}} \right)^{h/2} I_h\left(2\sqrt{\lambda_{1,n} \lambda_{2,n}}\right), \quad h \in \mathbb{Z},
\end{aligned}
\end{equation}
where $\mathbb{E}(Z_{n}) = \lambda_{1,n} - \lambda_{2,n}$
and $\text{Var}(Z_{n}) = \lambda_{1,n} + \lambda_{2,n}$. Furthermore,  $I_h(\cdot)$ is the modified Bessel function of order $h$ \citep{skellam1946frequency}.
The parameters $\lambda_{1,n}$ and $\lambda_{2,n}$ adopt
the same log-linear structure as in the previous cases. 


\subsection{Inflating the draws probability}
Poisson goal-based models often underestimate the incidence of draws, which are the diagonal elements in goal probability matrices. To mitigate this, \citet{karlis_ntzoufras2009} introduced a diagonally inflated bivariate Poisson (DIBP) model as follows
\begin{equation*}
\label{diag_poisson}
    \mathbb{P}_{X_{i,n},Y_{j,n}}(x_{i,n},y_{j,n})= \begin{cases}(1-\omega) \operatorname{BP}\left(\lambda_{1,n}, \lambda_{2.n}, \lambda_{3.n}\right) & \text { if } x_{i,n} \neq y_{j,n} \\ (1-\omega) \operatorname{BP}\left(\lambda_{1,n}, \lambda_{2,n}, \lambda_{3,n}\right)+\omega D(x_n, \xi) & \text { if } x_{i,n}=y_{j,n}\end{cases},
\end{equation*}
where $\operatorname{BP}(\cdot)$ is the bivariate Poisson probability mass function as in \eqref{eq:bivariate_poisson}, $\omega \in [0,1]$ controls the inflation weight, and $D(x_n, \xi)$ is a discrete distribution with parameter vector $\xi$, which favours draw outcomes.

Similarly, to address excess draws in goal differences, the zero-inflated Skellam model (ZISM) \citep{karlis_ntzoufras2009} can be adopted
\begin{equation*}
\label{diag_poisson}
    \mathbb{P}_{Z_n}(z_n)= \begin{cases}(1-\omega) \operatorname{SM}\left(\lambda_{1,n}, \lambda_{2,n}\right) & \text { if } z_n \neq 0 \\ (1-\omega) \operatorname{SM}\left(\lambda_{1,n}, \lambda_{2,n}\right)+\omega D(0, \xi) & \text { if } z_n=0\end{cases},
\end{equation*}
where $\operatorname{SM}(\cdot)$ is the Skellam probability mass function as in \eqref{skellam}, and $D(0, \xi)$ is a discrete distribution that places extra mass at zero.
\subsection{Dynamic prior distributions and identifiability constraints} \label{sec: owen_proposal}
A structural limitation in the previous models is the assumption of static team-specific parameters, namely, teams are assumed to have a constant performance over time, determined by attack and defence abilities $\beta^\text{att}$ and $\beta^\text{def}$, respectively. However, the performance of teams tends to be dynamic -- between seasons and even from week to week -- due to factors such as summer and winter transfer windows reshaping lineups, injuries benching key players, or midseason coaching changes because of unsatisfactory results.

Several approaches have been proposed to dynamically model team-specific abilities \citep[among others]{Rue_2000, owen_2011,Koopman_2015, koopman2019forecasting}. In particular, \citet{owen_2011} extended the static framework by introducing a discrete-time evolution for team-specific effects. Specifically, the evolution component is
specified as a random walk for both the attack and defence parameters by centering the effect of seasonal time $\tau$ on the lagged effect in $\tau-1$. This allows the attack and defence parameters to vary between seasons or weeks. Therefore, for each team $T_i$, where $i = 1, \ldots, N_T$, and each period $\tau$, where $\tau = 2,\ldots,\mathcal{T}$, the prior distributions for the attack and defence abilities are usually defined as follows
\begin{equation}
\label{owen_1}
 \begin{aligned}
& \beta^\text{att}_{i, \tau} \mid \beta^\text{att}_{i, \tau-1}, \sigma \sim \mathrm{N}\left(\beta^\text{att}_{i, \tau-1}, \dfrac{1}{\sigma}\right) \\
& \beta^\text{def}_{i, \tau} \mid \beta^\text{def}_{i, \tau-1} , \sigma\sim \mathrm{N}\left(\beta^\text{def}_{i, \tau-1}, \dfrac{1}{\sigma}\right).
\end{aligned}   
\end{equation}
While for the initial period $\tau=1$, the prior distributions are initialised as
\begin{equation}
\label{owen_2}
\begin{aligned}
& \beta^\text{att}_{i, 1} \mid \mu_\text{att} , \sigma \sim \mathrm{N}\left(\mu_{\operatorname{att}}, \dfrac{1}{\sigma}\right) \\
& \beta^\text{def}_{i, 1}\mid \mu_\text{def}, \sigma \sim \mathrm{N}\left(\mu_{\operatorname{def}}, \dfrac{1}{\sigma}\right),
\end{aligned}
\end{equation}
where $\mu_{\mathrm{att}}$ and $\mu_{\mathrm{def}}$ are the prior means for the initial attack and defence abilities, respectively, and $\sigma$ is the common evolution precision, assumed constant over time and identical between all teams and both team-specific abilities.
To ensure identifiability, a zero-sum constraint \citep{Baio_Blangiardo2010,owen_2011} on the random effects within each period is required
\begin{equation}
\label{zero_constraint}
\sum_{i=1}^{N_T} \beta^{\mathrm{att}}_{i,\tau} = 0,
\quad
\sum_{i=1}^{N_T} \beta^{\mathrm{def}}_{i,\tau} = 0,
\quad
\tau = 1,\dots,\mathcal{T}.
\end{equation}

As a matter of parameter interpretation, once the models have been estimated, a larger team-attack parameter indicates stronger attacking quality, while a smaller team-defence parameter corresponds to stronger defensive performance.

\section{A weighted dynamic proposal}\label{sec3}
As described in Section \ref{sec: owen_proposal}, a key assumption of the discrete‐time evolution approach as in \eqref{owen_1} is a single constant evolution precision $1/\sigma$ shared by all teams and by both their attack and defence parameters. However, this assumption can compromise predictive accuracy by either overborrowing (underborrowing) strength from one period to the next. Specifically, some periods -- such as the summer transfer window or a midseason coaching change -- can cause rapid shifts in team abilities, justifying the discount of earlier performance information; while other periods, when the teams' abilities are stable, borrowing more past information can improve the predictive performances. Furthermore, offensive and defensive abilities may often evolve at different rates. In this section, we propose a weighted dynamic approach based on commensurate priors, which employ separate, time‐varying evolution precisions for attack and defence. By treating the matches played during a specific period as "current data" with respect to the "historical data" from the previous period, this approach offers an intuitive framework in which the prior at each time point adaptively borrows information from the previous period, but only to the extent that the data justify it.

\subsection{Commensurate priors}
In the Bayesian framework, adaptively informative priors are valuable for synthesising results across studies, particularly in the clinical setting, where the appropriate borrowing of historical knowledge can be critical. By providing a coherent statistical framework that incorporates all relevant sources of information, these methods can substantially reduce the required sample sizes, increase statistical power, and lower both costs and ethical risks.

Here, we focus on hierarchical models that employ commensurate priors \citep{hobbs2011hierarchical, Hobbs2012} as the main mechanism to weight prior information according to its consistency (commensurability) with data from previous studies.
\citet{hobbs2011hierarchical} consider the case in which data from a single historical study inform the analysis of a new study by defining the commensurate prior for the parameter of interest $\theta$ as follows
\begin{equation}
\label{eqn: commensurate prior}
    \theta \mid \theta_0, \phi \sim \mathrm{N}\left(\theta_0, \frac{1}{\phi}\right),
\end{equation}
where $\theta_0$ is the estimate from the historical study and $\phi$ is the precision or commensurability parameter. 
The formulation in \eqref{eqn: commensurate prior} follows from the insight in \citet{pocock1976combination} for which historical parameters may be biased representations of their current counterparts. By modelling the unknown bias as $\epsilon = \theta-\theta_0$ the commensurate prior quantifies how much a current study parameter is allowed to vary with respect to the historical estimate in the absence of strong evidence of heterogeneity. Thus, a lack of evidence for substantial bias implies that the historical and current parameters are commensurate.

\citet{Hobbs2012} extended this framework by proposing both empirical and fully Bayesian methods to estimate or assign $\phi$. Notably, by incorporating prior uncertainty when estimating $\phi$, the fully Bayesian approach reduces the risk of overstating commensurability. \citet{Hobbs2012} proposed two families of priors for $\phi$, a family of gamma distributions that leads to a full conditional posterior distribution, as well as a variant of the “spike-and-slab” distribution introduced by \citet{mitchell1988bayesian} for Bayesian variable selection based on a mixture prior with
two components, which can provide robust borrowing. Specifically, the spike-and-slab prior distribution is a discrete mixture distribution
defined as locally uniform between two limits $0\leq\alpha_1<\alpha_2$ (the slab component), and with a probability mass concentrated at a point $\mathcal{S}>\alpha_2$ (the spike component), such that 
\begin{equation}\label{eq:spike-slab}
\begin{aligned}
    &\mathbb{P}(\phi<\alpha_1)=0, \\
&\mathbb{P}(\phi<u)=p_{l}\times\frac{u-\alpha_1}{\alpha_2-\alpha_1},\quad \alpha_1\le u\le \alpha_2,\\
&\mathbb{P}(\phi>\alpha_2)=\mathbb{P}(\phi=\mathcal{S})=1-p_{l},
\end{aligned}
\end{equation}
where $p_{l}$ is the probability of a slab, which can be interpreted as the prior probability of
incommensurability. The spike component concentrates the probability mass near $\theta_0$, encouraging strong borrowing from historical data, while the slab component allows for greater deviation when current data conflict with historical evidence. Thus, commensurate priors provide a mechanism for selectively borrowing information from historical data by using the adaptive shrinkage properties of spike-and-slab distributions. Notably, when the current and historical parameters appear commensurate, the prior strongly shrinks the current parameter towards the historical estimate, improving efficiency. Conversely, when there is substantial disagreement, the prior has minimal influence on $\theta$, limiting bias \citep{murray2015combining}. Indeed, with appropriate calibration, the spike-and-slab commensurate prior approach achieves desirable frequentist properties -- such as controlled Type I error and high power -- while adaptively borrowing information when it is commensurate and downweighting it when it is not \citep{Hobbs2012}.

\subsection{Weighted dynamic prior distributions}

 Let $\beta_{i,\tau}^{(k)}$ denote team $T_i$’s ability of type $k$ in period $\tau$, where $k \in \{\text{att}, \text{def}\}$ and $\tau = 1,2,\dots,\mathcal{T}$ indexes the time periods (e.g., seasons or weeks). In our weighted dynamic approach, each team’s ability in period $\tau$ has a prior centred on its ability from the previous period $\tau-1$, rather than assuming a fixed random walk precision across all periods. Therefore, for each team $T_i$, where $i = 1, \ldots, N_T$, and each period $\tau$, where $\tau = 2,\ldots,\mathcal{T}$, the prior distributions for the attack and defence abilities are
\begin{equation}\label{eq:weight_dyn_priors}
 \begin{aligned}
& \beta^\text{att}_{i, \tau} \mid \beta^\text{att}_{i, \tau-1}, \phi_{\text{att},\tau} \sim \mathrm{N}\left(\beta^\text{att}_{i, \tau-1}, \dfrac{1}{\phi_{\text{att},\tau}}\right) \\
& \beta^\text{def}_{i, \tau} \mid \beta^\text{def}_{i, \tau-1} , \phi_{\text{def},\tau} \sim  \mathrm{N}\left(\beta^\text{def}_{i, \tau-1}, \dfrac{1}{\phi_{\text{def},\tau}}\right),
\end{aligned}   
\end{equation} 
where each team’s offensive (defensive) ability in period $\tau$ has a normal prior distribution centred on the offensive (defensive) ability of that team in period $\tau-1$, with a commensurate (precision) parameter $\phi_{k,\tau}$ that governs how closely the agreement is with previous information at time $\tau-1$. If $\phi_{k,\tau}$ is large, the prior is tightly concentrated around the previous value – effectively assuming the team’s ability has not changed much – which leads to a heavy borrowing of strength from the previous period. Conversely, if $\phi_{k,\tau}$ is near zero, the prior is diffuse, indicating that we allow the current data to have a dominant influence while minimising the contribution of the previous data. Notably, we introduce separate precision $\phi_{\text{att},\tau}$ and $\phi_{\text{def},\tau}$ for each period $\tau$ and, similarly to \citet{egidi_etal2018}, for each ability type, rather than a common evolution precision. This means that the model can adjust how much it learns from the attack strength of the previous period independently of how much it learns from the defence strength of the previous period.

To complete the model specification, we assign spike-and-slab hyperpriors to each precision parameter. Rather than using a discrete spike-and-slab with a point mass at $\mathcal{S}$ and a uniform slab on $[\alpha_1,\alpha_2]$ as in \eqref{eq:spike-slab}, we employ a continuous two-component mixture consisting of a highly concentrated spike and a diffuse slab \citep{hong2018power}. For each period and ability of type $k$, where $k \in \{\text{att}, \text{def}\}$, we let
\begin{equation} \label{eq:weight_dyn_hyperpriors}
    \begin{aligned}
        \phi_{k,\tau} \mid \mu_{s}, \mu_{l},\psi_{s},\psi_{l},p_{l}  & \sim  \mathrm{N}^+(\mu_{s},\psi_{s}) \times (1 - p_{l}) + \mathrm{N}^+(\mu_{l},\psi_{l})\times p_{l} ,
    \end{aligned}
\end{equation}
where $\mathrm{N}^+(\mu,\psi)$ denotes a normal distribution with mean $\mu$ and standard deviation $\psi$ truncated from below at zero (i.e., the half-normal distribution). Specifically, $\mu_s$ and $\mu_l$ represent the means of the spike-and-slab components, respectively, while $\psi_s$ and $\psi_l$ are the corresponding standard deviations, with $0 < \psi_s < \psi_l$. Each $\phi_{k,\tau}$, with $\tau \geq2$, thus has a chance to be larger -- implying a high commensurability with the previous period -- or to be near zero -- suggesting low commensurability and allowing more variability with respect to the previous period. The model will estimate an appropriate value for $\phi_{\text{att},\tau}$ and $\phi_{\text{def},\tau}$ based on the degree to which the new match results align with the trend of the previous period. If the performance of the teams in period $\tau$ looks very similar to that of period $\tau-1$, the posterior for $\phi_{k,\tau}$ will likely favour the spike, implying strong borrowing and shrinkage towards the past. In contrast, if the performance of the teams changes unexpectedly, the posterior of $\phi_{k,\tau}$ will move toward the slab, implying weak borrowing.
For the initial period $\tau=1$, no past information is available, so we use diffuse but proper priors as in \eqref{owen_2}. Furthermore, to ensure the identifiability of the model, we impose the zero sum constraints of each period as in \eqref{zero_constraint}.

Our weighted dynamic approach extends the Bayesian dynamic goal-based models framework by introducing adaptive period-specific shrinkage for team abilities. By allowing the data to decide how much information about attack and defence abilities to borrow from the previous period, the model can reflect real-world changes more responsively. This yields a more flexible evolution of team strengths over time, which should improve predictive performance on match outcomes.

\section{Application}\label{sec4}
We evaluate the efficacy of our proposed model using six Bayesian dynamic goal-based models, as described in Section \ref{sec2}. Our analysis uses data from the five most recent seasons (2020/2021 through 2024/2025) of three major European football leagues: the German Bundesliga , the English Premier League (EPL), and the Spanish La Liga.
We compare the performance of our proposal with the discrete-time approach of \citet{owen_2011}, which uses a single evolution precision shared between all teams and for both offensive and defensive abilities. Additionally, we include the extension by \citet{egidi_etal2018}, which introduces a constant evolution precision that is specific to either attack or defence abilities but shared among all teams.
Each season in our datasets is treated as two discrete-time periods -- the first half and the second half of the season, resulting in a total of ten time periods for analysis. This division is designed to capture midseason structural events, such as winter transfer windows and breaks, which can significantly alter team composition and performance. For instance, midseason often brings roster changes (through transfers), managerial turnovers, and player recovery from injuries, all of which can change the performance of a team in the last half of the season. To comprehensively assess predictive performance, we consider three distinct prediction scenarios for each league: the entire second half, the last three rounds, and the last round of the most recent season. These scenarios cover different time horizons and levels of volatility, allowing us to examine how well each model adapts to changing conditions. In particular, forecasting over an entire half-season provides a broad view that incorporates the cumulative impact of all post-midseason changes (e.g., transfers, coaching changes, tactical adjustments). In contrast, focussing on the final three rounds focusses on a crucial segment of the season often marked by intensified competitive pressure -- during this phase, results can decide championships, European qualification, or relegation, and teams may adjust their strategies accordingly (e.g., rotating squads to manage fatigue, or adopting more aggressive or defensive tactics as needed). Finally, predicting only the last round represents the most uncertain scenario. In the final round, teams have widely varying motivations -- some are competing for crucial objectives, while others have little or nothing to lose -- which often leads to surprising outcomes. By examining performance in these three scenarios, we can assess the robustness and adaptability of our weighted dynamic models under a range of realistic competitive conditions.

The models are implemented using the probabilistic programming language Stan \citep{stan}, employing Markov Chain Monte Carlo (MCMC) sampling via the R package \texttt{footBayes} \citep{footBayes}. For posterior sampling, we run four independent chains, each consisting of 2000 iterations, with the initial 1000 iterations discarded as burn-in. In the spike-and-slab formulation as in \eqref{eq:weight_dyn_hyperpriors}, the spike component has mean $\mu_s = 100$ and standard deviation $\psi_s = 0.1$, while the slab component has mean $\mu_l = 0$ and standard deviation $\psi_l = 5$, that is,
\begin{equation*} 
    \begin{aligned}
        \phi_{\text{att},\tau}, \phi_{\text{def},\tau} \mid p_l & \sim  \mathrm{N}^+(100,0.1) \times (1-p_l) + \mathrm{N}^+(0,5)\times p_l,
    \end{aligned}
\end{equation*}
where $p_l$ is the prior probability of drawing from the slab that is set to $0.99$ \citep{Hobbs2012,chen2018web}. The chosen half-normal slab prior is approximately uniform over
$[0, 3]$ and decays afterwards, allowing either minimal borrowing or complete discounting of prior information when necessary \citep{alt2025hdbayes}. In contrast, the spike prior is structured to yield complete pooling when the past information aligns closely with current observations \citep{ouma2022bayesian, zheng2022borrowing, chen2018web, Hobbs2012}.
For the approaches proposed by \citet{owen_2011} and \citet{egidi_etal2018}, the evolution precisions are modelled as
\begin{align*}
\sigma, \sigma_{\text{att}}, \sigma_{\text{def}} \sim \text{Cauchy}^+(0,5),
\end{align*}
where $\text{Cauchy}^+(0,\nu)$ denotes the half-Cauchy distribution with location 0 and scale $\nu$. As noted in \citet{Gelman2006Prior}, the half-Cauchy distribution is a flexible and weakly informative prior for scale parameters in hierarchical models, with advantageous behaviour near zero and minimal influence on posterior estimates. Finally, for all models, a weakly informative prior \citep{gelman2008weakly} is assigned to the home-effect parameter
\begin{align*}
    \text{home}\sim \mathrm{N}(0,5).
\end{align*}

\subsection{Predictive performance}
\label{sub_sec_pred_perf}

One of the key aspects of sports analytics is the ability to generate accurate future predictions. Bayesian models naturally provide posterior probabilities for future matches. Considering the posterior predictive distribution for future observable data $\tilde{\mathcal{D}}$, we incorporate the predictive uncertainty of the model, propagated from the uncertainty of the posterior parameter. Predictions are generated by conditioning future observable values on the posterior parameter estimates
\begin{align*}
    p(\tilde{\mathcal{D}}| \mathcal{D}) = \int p(\tilde{\mathcal{D}}| \boldsymbol{\theta}) \pi(\boldsymbol{\theta}| \mathcal{D}) d\boldsymbol{\theta}.
\end{align*}
After obtaining predictions from the models, evaluating their performance is crucial to assessing their predictive power and reliability. Specifically, we focus on two predictive metrics to rigorously examine the predictive performance of the models described in Section \ref{sec2}. Additional analyses with two other predictive metrics are provided in Appendix \ref{secA1}.

\subsubsection{Predictive metrics}
\label{sec_brier_RPS}
The evaluation of probabilistic forecasts typically involves scoring rules metrics, which assess forecast performance by comparing predictions with the corresponding outcomes.
The Brier score \citep{brier1950verification}, recommended by \citet{spiegelhalter2009one}, is a non-local and distance-insensitive scoring rule, essentially acting as a mean squared error for forecasts. It is defined as
\begin{equation*}
    \text{Brier} = \frac{1}{M}\sum_{m=1}^M\sum_{r=1}^3(p_{r,m}-\delta_{r,m})^2,
\end{equation*}
where $p_{r,m}$ denotes the predicted probability of outcome $r$, with $r \in \{\text{home win, draw, away win}\}$, for the $m$-th match played during the forecast period. Here, $\delta_{r,m}$ is the Kronecker delta, that is, $1$ if the outcome $r$ occurs in the $m$-th match. The Brier score ranges from $0$, indicating perfect prediction accuracy, to a maximum of $2$ when predictions consistently assign probability $1$ to incorrect outcomes.

While proper scoring rules penalise squared errors, mean-based metrics provide direct, interpretable summaries of predictive accuracy. The Average of Correct Probabilities (ACP) is defined as the arithmetic mean of the probabilities assigned to outcomes that actually occurred, that is
\begin{align*}
    \text{ACP} = \frac{1}{M}\sum_{m = 1}^M p_{o, m},
\end{align*}
where $p_{o, m}$ is the probability assigned to the observed outcome of the $m$-th match. Being an arithmetic mean, the ACP measures the confidence of the average forecast directly on the original probability scale. ACP values near 1 indicate that the model consistently assigns high probabilities to the true outcomes, whereas values near 0 reflect weaker predictive performance. 

\begin{figure}[htb!]
    \centering   \includegraphics[width=0.99\textwidth, height=11.5cm]{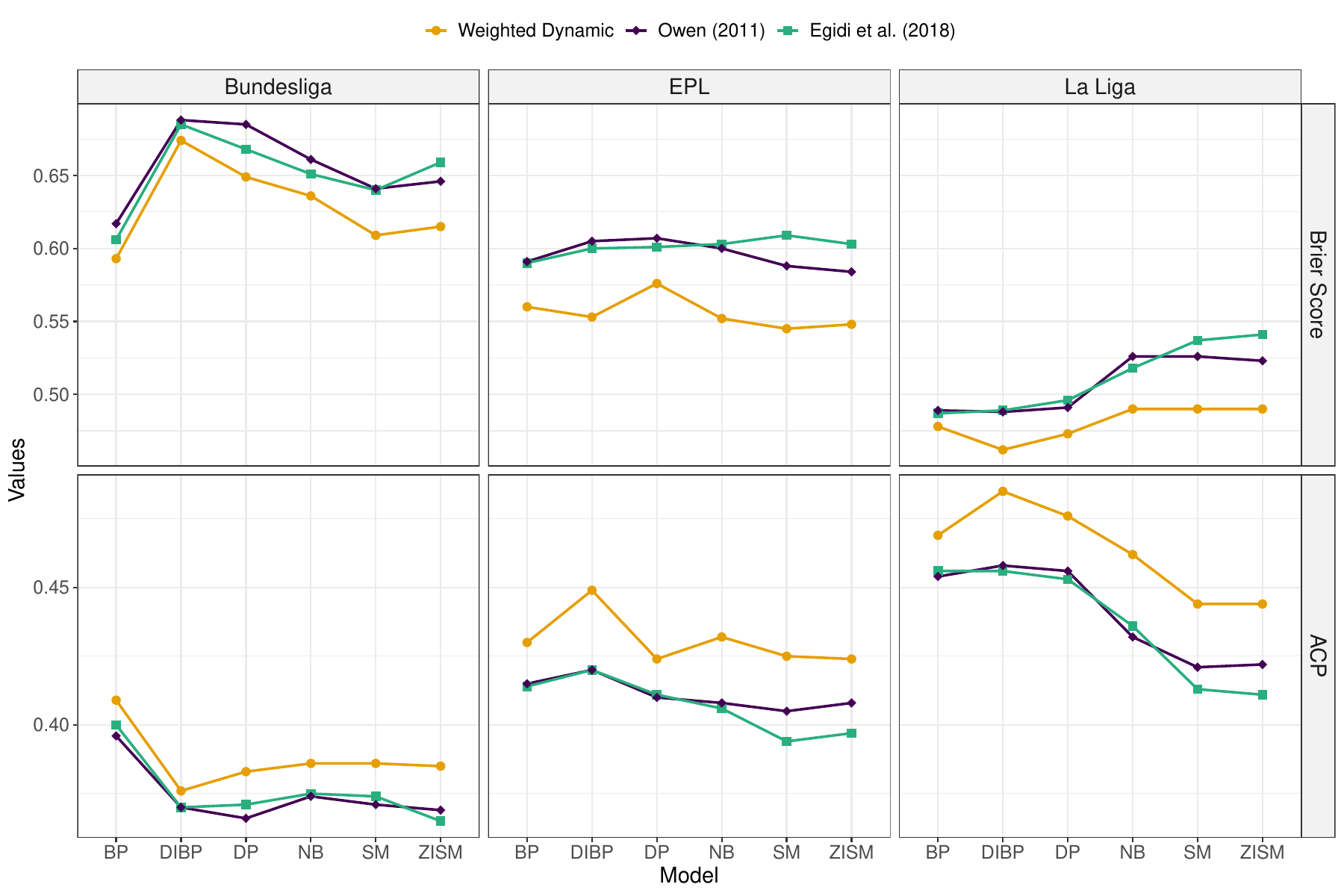}
        \caption{Lineplot comparing Brier Score and Average of Correct Probabilities (ACP) for the proposed weighted dynamic method with those of \citet{owen_2011} and \citet{egidi_etal2018}, evaluated on the final round of the 2024/2025 season. The comparison includes six models: Bivariate Poisson (BP), Diagonal-Inflated Bivariate Poisson (DIBP), Double Poisson (DP), Negative Binomial (NB), Skellam Model (SM), and Zero-Inflated Skellam Model (ZISM), for the Bundesliga, La Liga, and English Premier League (EPL).}
        \label{brier}
\end{figure}

Figure~\ref{brier} compares the Brier score and the ACP for the proposed weighted dynamic approach compared to those of \citet{owen_2011} and \citet{egidi_etal2018}, evaluated in the final round of the 2024/2025 season scenario. The plot includes the bivariate Poisson, diagonal-inflated bivariate Poisson, double Poisson, negative binomial, Skellam model, and zero-inflated Skellam model. The proposed weighted dynamic approach consistently demonstrates superior predictive performance, yielding the lowest values for the Brier score and the highest values for the ACP among all models and competitions. In the Bundesliga, the bivariate Poisson model achieves the smallest Brier score with a value of $0.593$ and the largest ACP value of $0.409$. For the EPL, the Skellam model obtains the lowest Brier score at $0.545$ while the diagonal-inflated bivariate Poisson reaches an ACP of $0.449$. In La Liga, the diagonal-inflated bivariate Poisson achieves a Brier score of $0.462$ and an ACP of $0.485$.

Table~\ref{tab_1} summarises the Brier scores and ACPs for each of the six goal‐based models, comparing our weighted‐dynamic forecasts with those of \citet{owen_2011} and \citet{egidi_etal2018} over the last three matchdays of the 2024/25 season. 
Among the three leagues, the weighted dynamic approach consistently achieves the lowest Brier score and the highest ACP values, reflecting more accurate predictions on decisive matches at the end of the season.
In the Bundesliga, our weighted dynamic approach reduces the Brier score from $0.683$ to $0.678$ compared to the method proposed by \citet{egidi_etal2018} in the bivariate Poisson model and presents the highest ACP value of $0.359$. In the EPL, the weighted dynamic models outperform the other approaches, yielding the lowest and highest values for the Brier score and the ACP, respectively. Notably, in the diagonal-inflated bivariate Poisson model, the Brier score decreases to $0.602$, while the ACP reaches a value of $0.421$, highlighting the benefit of draw inflation combined with period‐specific weighting in a highly unpredictable competition. Similarly, in La Liga, the weighted-dynamic models outperform all other approaches, achieving the best results with a Brier score of $0.499$ under the diagonal-inflated bivariate Poisson model, while showing the highest ACP of $0.454$.
\begin{table}[h!] 
\caption{Brier Score and Average of Correct Probabilities (ACP) for the proposed weighted dynamic method, \citet{owen_2011} method and \citet{egidi_etal2018} method, evaluated on the last three rounds of the 2024/2025 season for the Bundesliga, English Premier League (EPL), and La Liga.}
\label{tab_1}
\centering
\begin{adjustbox}{max width=\textwidth}
\begin{tabular}{l
                l
                S[table-format=1.4, table-space-text-pre={\B{0}}] S[table-format=1.4, table-space-text-pre={\B{0}}]
                S[table-format=1.4, table-space-text-pre={\B{0}}] S[table-format=1.4, table-space-text-pre={\B{0}}]
                S[table-format=1.4, table-space-text-pre={\B{0}}] S[table-format=1.4, table-space-text-pre={\B{0}}]}
\toprule
\textbf{League} & \textbf{Model} & 
\multicolumn{2}{c}{\textbf{Weighted Dynamic}} & 
\multicolumn{2}{c}{\textbf{Owen (2011)}} & 
\multicolumn{2}{c}{\textbf{Egidi et al. (2018)}} \\
& & \textbf{Brier Score} & \textbf{ACP} & \textbf{Brier Score} & \textbf{ACP} & \textbf{Brier Score} & \textbf{ACP} \\
\midrule
Bundesliga & Bivariate Poisson     & \B{0.678} & \B{0.359} & 0.687 & 0.357 & 0.683 & 0.357 \\
 & Diag. Infl. Bivariate Poisson  & 0.735 & 0.341 & 0.733 & 0.344 & 0.725 & 0.348 \\
 & Double Poisson   & 0.718 & 0.344 & 0.719 & 0.346 & 0.702 & 0.353 \\
 & Negative Binomial & 0.698 & 0.351 & 0.706 & 0.350 & 0.690 & 0.357 \\
 & Skellam Model & 0.688 & 0.345 & 0.688 & 0.347 & 0.684 & 0.351 \\
 & Zero Infl. Skellam Model & 0.693 & 0.344 & 0.690 & 0.348 & 0.686 & 0.351 \\
 \midrule
EPL & Bivariate Poisson     & 0.603 & 0.409 & 0.608 & 0.402 & 0.609 & 0.402 \\
 & Diag. Infl. Bivariate Poisson  & \B{0.602} & \B{0.421} & 0.616 & 0.410 & 0.619 & 0.408 \\
 & Double Poisson   & 0.612 & 0.408 & 0.622 & 0.401 & 0.626 & 0.399 \\
 & Negative Binomial & 0.606 & 0.410 & 0.612 & 0.404 & 0.611 & 0.403 \\
 & Skellam Model & 0.617 & 0.393 & 0.624 & 0.387 & 0.617 & 0.389 \\
 & Zero Infl. Skellam Model & 0.615 & 0.394 & 0.624 & 0.387 & 0.624 & 0.385 \\
\midrule
La Liga & Bivariate Poisson     & 0.502 & 0.448 & 0.520 & 0.438 & 0.518 & 0.439 \\
 & Diag. Infl. Bivariate Poisson  & \B{0.499} & \B{0.454} & 0.518 & 0.444 & 0.521 & 0.442 \\
 & Double Poisson   & 0.503 & 0.450 & 0.522 & 0.439 & 0.518 & 0.441 \\
 & Negative Binomial & 0.514 & 0.442 & 0.533 & 0.431 & 0.534 & 0.430 \\
 & Skellam Model & 0.553 & 0.408 & 0.554 & 0.407 & 0.556 & 0.406 \\
 & Zero Infl. Skellam Model & 0.548 & 0.411 & 0.554 & 0.407 & 0.555 & 0.407 \\
\bottomrule
\end{tabular}
\end{adjustbox}
\end{table}

Table~\ref{tab_2} presents the same comparisons for the second half of last season. Specifically, in the Bundesliga, the weighted dynamic bivariate Poisson model achieves the lowest Brier score of $0.661$ and the highest ACP value of $0.387$. In the EPL, the weighted dynamic bivariate Poisson obtains the lowest Brier Score of $0.579$, while the weighted dynamic diagonal-inflated bivariate Poisson model produces an ACP of $0.429$. Similarly, in La Liga, the weighted dynamic diagonal-inflated bivariate Poisson model achieves a Brier score of $0.583$ and produces the highest ACP $(0.425)$.
\begin{table}[htb!]
\caption{Brier Score and Average of Correct Probabilities (ACP) for the proposed weighted dynamic method, \citet{owen_2011} method and \citet{egidi_etal2018} method, evaluated on the second half of the 2024/2025 season for the Bundesliga, English Premier League (EPL), and La Liga.}
\label{tab_2}
\centering
\begin{adjustbox}{max width=\textwidth}
\begin{tabular}{l
                l
                S[table-format=1.4, table-space-text-pre={\B{0}}] S[table-format=1.4, table-space-text-pre={\B{0}}]
                S[table-format=1.4, table-space-text-pre={\B{0}}] S[table-format=1.4, table-space-text-pre={\B{0}}]
                S[table-format=1.4, table-space-text-pre={\B{0}}] S[table-format=1.4, table-space-text-pre={\B{0}}]}
\toprule
\textbf{League} & \textbf{Model} & 
\multicolumn{2}{c}{\textbf{Weighted Dynamic}} & 
\multicolumn{2}{c}{\textbf{Owen (2011)}} & 
\multicolumn{2}{c}{\textbf{Egidi et al. (2018)}} \\
& & \textbf{Brier Score} & \textbf{ACP} & \textbf{Brier Score} & \textbf{ACP} & \textbf{Brier Score} & \textbf{ACP} \\
\midrule
Bundesliga & Bivariate Poisson     & \B{0.661} & \B{0.387} & 0.664 & 0.382 & 0.662 & 0.384 \\
 & Diag. Infl. Bivariate Poisson  & 0.694 & 0.386 & 0.691 & 0.383 & 0.694 & 0.383 \\
 & Double Poisson   & 0.677 & 0.386 & 0.683 & 0.381 & 0.685 & 0.380 \\
 & Negative Binomial & 0.671 & 0.384 & 0.680 & 0.378 & 0.682 & 0.377 \\
 & Skellam Model & 0.664 & 0.370 & 0.668 & 0.368 & 0.667 & 0.370 \\
 & Zero Infl. Skellam Model & 0.664 & 0.372 & 0.668 & 0.369 & 0.667 & 0.371 \\
 \midrule
EPL & Bivariate Poisson     & \B{0.579} & 0.422 & 0.581 & 0.421 & 0.583 & 0.420 \\
 & Diag. Infl. Bivariate Poisson  & 0.594 & \B{0.429} & 0.592 & 0.427 & 0.588 & 0.428 \\
 & Double Poisson   & 0.584 & 0.424 & 0.584 & 0.424 & 0.582 & 0.425 \\
 & Negative Binomial & 0.584 & 0.421 & 0.582 & 0.422 & 0.583 & 0.422 \\
 & Skellam Model & 0.601 & 0.399 & 0.600 & 0.399 & 0.597 & 0.401 \\
 & Zero Infl. Skellam Model & 0.604 & 0.398 & 0.597 & 0.401 & 0.596 & 0.402 \\
\midrule
La Liga & Bivariate Poisson     & 0.583 & 0.416 & 0.586 & 0.413 & 
0.585 & 0.414 \\
 & Diag. Infl. Bivariate Poisson  & \B{0.583} & \B{0.425} & 0.586 & 0.421 & 0.586 & 0.421 \\
 & Double Poisson   & 0.585 & 0.419 & 0.585 & 0.416 & 0.585 & 0.416 \\
 & Negative Binomial & 0.590 & 0.415 & 0.591 & 0.411 & 0.587 & 0.413 \\
 & Skellam Model & 0.583 & 0.401 & 0.589 & 0.395 & 0.594 & 0.394 \\
 & Zero Infl. Skellam Model & 0.583 & 0.402 & 0.588 & 0.396 & 0.592 & 0.396 \\
\bottomrule
\end{tabular}
\end{adjustbox}
\end{table}

\subsection{Team abilities}
\begin{figure}[htb!]
    \centering   \includegraphics[width=0.99\textwidth, height=12cm]{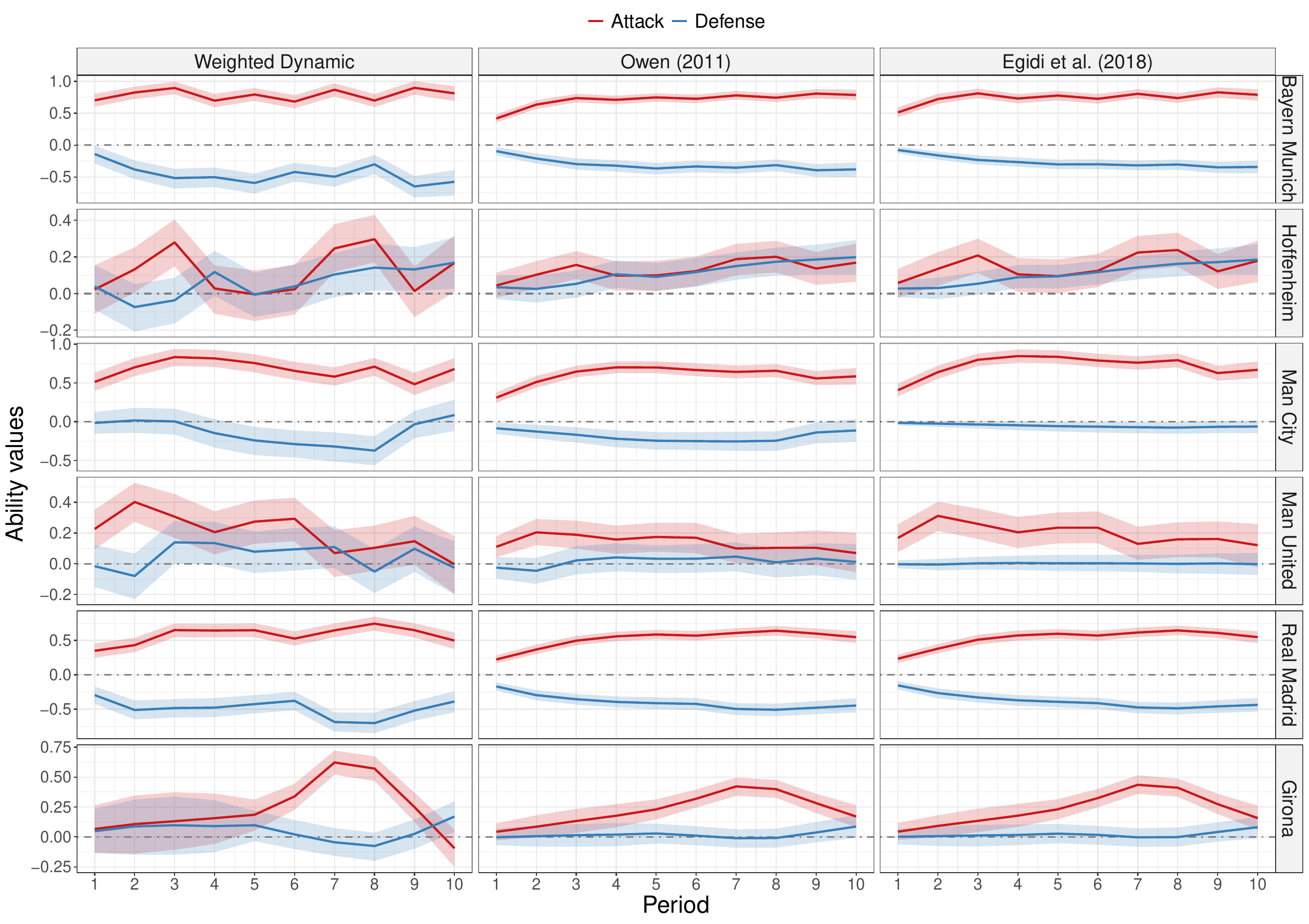}
\caption{Trajectories of estimated attacking (solid red lines) and defensive (solid blue lines) abilities with their 50\% credible intervals over ten periods for two representative teams in each league.  Results from the proposed weighted dynamic method are shown alongside corresponding estimates from \citet{owen_2011} and \citet{egidi_etal2018}, all evaluated at the final round of the 2024/2025 season scenario.}
\label{ability_plot}
\end{figure}

One of the crucial aspects of our proposal is how the weighted dynamic approach in \eqref{eq:weight_dyn_priors} influences the evolution of the attacking and defensive abilities of the teams in the evaluated periods. Figure~\ref{ability_plot} illustrates the trajectories of these abilities for the best performing model identified in Section~\ref{sub_sec_pred_perf}, when forecasting the final round of the 2024/2025 season.  Specifically, for the Bundesliga the best model is the bivariate Poisson model, for the EPL the Skellam model, and for La Liga the diagonal-inflated bivariate Poisson model. Within each league, we selected two teams: one with relatively stable performance during the study period and another showing inconsistent behaviour. The plot shows that, for stable teams, the weighted dynamic model captures the temporal trends more accurately, with clearer distinctions between attacking and defensive strength. For instance, Bayern München, typically dominant in the Bundesliga, showed a slight slump in performance during the 2023/2024 season (periods 7 and 8) when they finished third. This fluctuation is reflected in the attack and defence abilities, particularly in the weighted dynamic model.
Similarly, for Real Madrid and Manchester City, both of which won their respective leagues in 2023/2024 but finished second and third in 2024/2025 (periods 9 and 10), the weighted dynamic approach captures a noticeable drop in attacking ability and a relative increase in defensive ability. These changes are more distinctly represented in our model compared to the alternatives of \citet{owen_2011} and \citet{egidi_etal2018}.

The benefits of adaptive shrinkage are even more evident for teams with inconsistent performance. For instance, Manchester United, after finishing second in 2020/2021, showed a gradual decline in subsequent seasons, finally placing 15th in the 2024/2025 season. After yielding the highest and lowest values for the attack and defence abilities, respectively, compared to the other two proposals for the 2020/2021 season, the weighted dynamic model reflects this trend with a notable reduction in attacking ability and an increase in defensive vulnerability, particularly pronounced in the final periods.
Similarly, Girona FC experienced an exceptional 2023/2024 season, reaching a third place in La Liga, followed by a disappointing 16th place finish in 2024/2025. The weighted dynamic approach effectively captures this volatility, showing the highest attacking ability during the best season (periods 7 and 8) and a marked decline thereafter. The model even shows a crossover point in the final period, where Girona's defensive vulnerability is higher than its attacking ability, highlighting a shift in team abilities that other models fail to capture.

Additional analyses on the commensurate precisions for the offensive and defensive abilities parameters are provided in the Appendix \ref{app_comm_par}.

\section{Discussion}\label{sec5}
This work introduced a Bayesian weighted dynamic framework for football predictions that flexibly models the evolution of team-specific abilities over time. By using commensurate priors with spike-and-slab hyperparameters, our approach allows each team's attack and defence strength in a given period to adaptively borrow information from past performance. This yields a more responsive and nuanced dynamic model compared to previous approaches that assume static abilities or a single constant evolution precision. Among six different goal-based distributions and three major European leagues, we found that this adaptive shrinkage mechanism leads to consistent improvements in predictive accuracy and a substantial reduction in computational time relative to earlier dynamic models. The weighted dynamic models captured team performance trajectories more realistically – for instance, they sharply reflected midseason form fluctuations and major transitions – while maintaining or improving forecast prediction accuracy. The greatest gains emerged in short-term prediction tasks (e.g., the final rounds of a season), where the ability to adjust quickly to recent surprises is crucial. Furthermore, in terms of computation time, the proposed weighted dynamic approach outperforms the dynamic formulations of
\citet{owen_2011} and \citet{egidi_etal2018} while achieving satisfactory convergence diagnostics for all scenarios evaluated. Specifically, among all leagues and models, the weighted dynamic model
consistently requires the least computation time to reach convergence. More details are provided in Appendix \ref{app_comp_conv}.

In this paper, we maintained the dependence parameter in the bivariate Poisson models constant, but future work could investigate making the covariance of scores dynamic if needed. These additions would offer a comprehensive overview of how every aspect of the game evolves, albeit at the cost of $\mathcal{T}-1$ additional parameters. Furthermore, incorporating additional predictors or team covariates may improve predictions. Our current implementation used only past match results to infer team abilities; however, information such as team market value, recent injuries, or even in-game statistics (shots, expected goals) could be included as covariates influencing the scoring rates parameters. Future extensions might also consider team-specific or hierarchical evolution parameters, so that traditionally inconsistent teams are allowed more variation than stable teams.

Another important direction is the integration with result-based models and other comparative approaches. Although we focused on goal-based distributions (which naturally yield the three-way process as a consequence), our methodology could be extended to models that predict match outcomes directly. For instance, in an ordered probit/logit model or a multi-class logistic model for win–draw–loss, one could let each team’s latent strength parameter vary over time using the same weighted random-walk prior. 
In addition, a weighted dynamic Bradley–Terry-Davidson model for paired comparisons is a natural extension. Our approach could provide a fully Bayesian weighted dynamic Bradley–Terry-Davidson model by allowing each team’s strength to evolve with our weighted dynamic approach. Conceptually, this would let the probability of one team beating another adapt rapidly after major changes (e.g., if a traditionally weak team suddenly improves, the model would downweight its past information). We expect that such a model would be computationally even simpler (since it has only one strength per team rather than separate attack/defence), yet still benefit from our weighted dynamic approach.


We emphasise that the Bayesian weighted dynamic approach presented here is quite general and may find use in other sports domains. Many sports and competitive systems (e.g., basketball, volleyball, or handball) involve teams whose skills change over time. By calibrating the commensurate prior to the specific domain, our strategy of time-specific shrinkage could be applied wherever one has sequential performance data and expects occasional shifts in the underlying ability.  Furthermore, while our focus here has been on domestic leagues with complete round‑robin schedules, future research will focus on high‑profile tournaments feature group stages followed by knockout rounds or hybrid formats such as the UEFA Champions League, FIFA World Cup, and UEFA European Championship. Finally, the proposed method is implemented in the free and open source R package \texttt{footBayes}.

\section*{Software and Data Availability}
All analyses were performed in the R programming language version $4.4.3$ \citep{R_software_2023}. Data are freely available online at \href{https://www.football-data.co.uk/}{football-data.co.uk}. All computational simulations in Appendix \ref{app_comp_conv} were conducted on an Intel Core i7-1260P laptop with 16GB of RAM running Ubuntu 22.04. The code for reproducing this manuscript is openly available at \url{https://github.com/RoMaD-96/BayesWDFM}.  The proposed methodology has also been implemented in the free and open source R package \href{https://github.com/LeoEgidi/footBayes}{\texttt{footBayes}} (from version 2.1.0).
\section*{Acknowledgments}

This work has been supported by the project "SMARTsports: “Statistical Models and AlgoRiThms in sports. Applications in professional and amateur contexts, with able-bodied and disabled athletes”, funded by the MIUR Progetti di Ricerca di Rilevante Interesse Nazionale (PRIN) Bando 2022 - grant n. 2022R74PLE (CUP J53D23003860006).

\begin{appendices}

\section{Ranked Probability Score and $\text{Pseudo-R}^2$}\label{secA1}

For discrete outcomes, it can be beneficial for a scoring rule to account for the proximity or ordering of potential outcomes. In football, a draw is closer to a home win than an away win. The Ranked Probability Score (RPS) \citep{epstein1969scoring} is a distance-sensitive scoring measure that evaluates the degree to which the forecast probability distribution matches the observed outcome, assigning higher scores to probabilistic forecasts that allocate higher probabilities to outcomes near the actual result. The RPS is defined as
\begin{align*}
    \text{RPS} = \frac{1}{3-1}\sum_{r = 1}^{3-1}\left(\sum_{l =1}^r p_{l,m}-\sum_{l =1}^r\delta_{l,m}\right)^2.
\end{align*}
As with the Brier score, lower values indicate better predictive performance.

The $\text{pseudo-R}^2$ \citep{dobson2001economics} is
defined as the geometric mean of the probabilities assigned to the actual result of each match.
\begin{align*}
\text{Pseudo-R}^2 = \Bigl(\prod_{m = 1}^M p_{o, m}\Bigr)^{1/M}.
\end{align*}
The geometric mean penalises low-probability predictions more severely than the arithmetic mean. Similarly to ACP, a $\text{pseudo-R}^2$ close to 1 indicates high predictive accuracy, while values approaching 0 suggest weaker performance.

\begin{figure}[htb!]
    \centering   \includegraphics[width=0.99\textwidth, height=11.5cm]{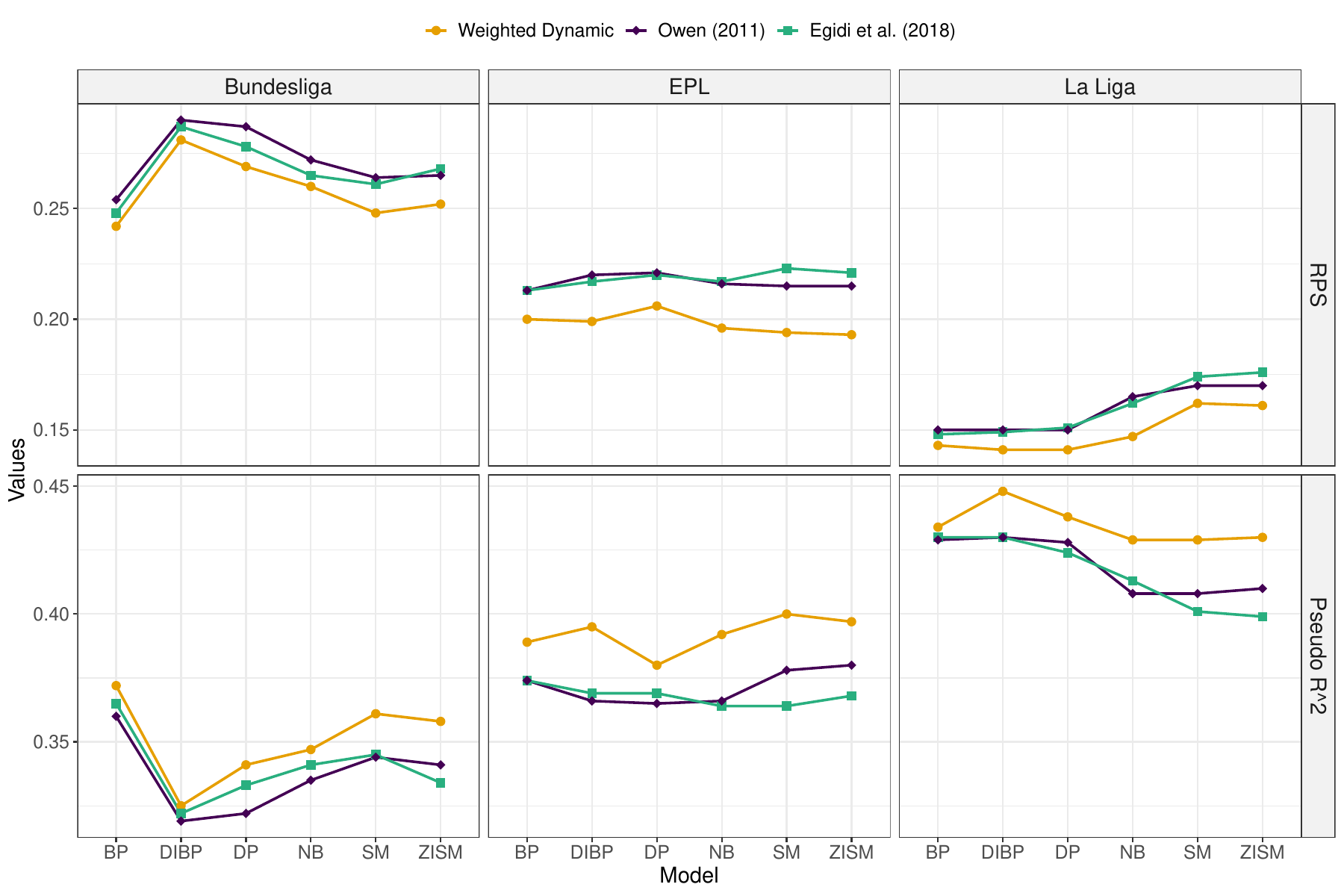}
        \caption{Lineplot comparing the Ranking Probability Score (RPS) and $\text{Pseudo-R}^2$ for the proposed weighted dynamic method with those of \citet{owen_2011} and \citet{egidi_etal2018}, evaluated on the final round of the 2024/2025 season. The comparison includes six models: Bivariate Poisson (BP), Diagonal-Inflated Bivariate Poisson (DIBP), Double Poisson (DP), Negative Binomial (NB), Skellam Model (SM), and Zero-Inflated Skellam Model (ZISM), for the Bundesliga, English Premier League (EPL), and La Liga.}
        \label{RPS}
\end{figure}

Figure~\ref{RPS} compares the RPS and the $\text{pseudo-R}^2$ for the proposed weighted dynamic approach compared to those of \citet{owen_2011} and \citet{egidi_etal2018}, evaluated in the final round of the 2024/2025 season. The proposed weighted dynamic approach consistently demonstrates superior predictive performance, yielding the lowest RPS and the highest $\text{pseudo-R}^2$ values among all models and competitions. In the Bundesliga, the bivariate Poisson model achieves the lowest RPS with a value of $0.242$ and a $\text{pseudo-R}^2$ of $0.372$. For the EPL, the zero-inflated Skellam model obtains the smallest RPS at $0.193$, while the Skellam model shows the highest $\text{pseudo-R}^2$ at $0.400$. In La Liga, the diagonal-inflated
bivariate Poisson model presents an RPS of $0.141$ and achieves the highest $\text{pseudo-R}^2$ at $0.448$.

Table~\ref{tab_rps_pseudoR_three_matches} presents the RPS and $\text{Pseudo-R}^2$ values for each of the six goal‐based models, comparing our weighted‐dynamic forecasts with those of \citet{owen_2011} and \citet{egidi_etal2018} over the last three matchdays of the 2024/25 season.  In the Bundesliga, our weighted dynamic approach slightly decreases the RPS from $0.217$ to $0.216$ compared to the method proposed by \citet{egidi_etal2018} while reaching the highest value for the $\text{Pseudo-R}^2$ (0.328) in the bivariate Poisson model. In the EPL, the weighted dynamic models yield the smallest and largest values for the RPS and $\text{Pseudo-R}^2$, respectively. Notably, in the diagonal-inflated bivariate Poisson model the RPS is $0.224$, while in the bivariate Poisson model the $\text{Pseudo-R}^2$ is $0.370$. Similarly, in La Liga, the weighted-dynamic models outperform all other approaches, achieving the best results with an RPS of $0.189$ and a $\text{Pseudo-R}^2$ of $0.421$ under the diagonal-inflated bivariate Poisson model.
\begin{table}[h!] 
\caption{Ranked Probability Score (RPS) and $\text{Pseudo-R}^2$ for the proposed weighted dynamic method, \citet{owen_2011} method and \citet{egidi_etal2018} method, evaluated on the last three round of the 2024/2025 season for the Bundesliga, English Premier League (EPL), and La Liga.}
\label{tab_rps_pseudoR_three_matches}
\centering
\begin{adjustbox}{max width=\textwidth}
\begin{tabular}{l
                l
                S[table-format=1.4, table-space-text-pre={\B{0}}] S[table-format=1.4, table-space-text-pre={\B{0}}]
                S[table-format=1.4, table-space-text-pre={\B{0}}] S[table-format=1.4, table-space-text-pre={\B{0}}]
                S[table-format=1.4, table-space-text-pre={\B{0}}] S[table-format=1.4, table-space-text-pre={\B{0}}]}
\toprule
\textbf{League} & \textbf{Model} & 
\multicolumn{2}{c}{\textbf{Weighted Dynamic}} & 
\multicolumn{2}{c}{\textbf{Owen (2011)}} & 
\multicolumn{2}{c}{\textbf{Egidi et al. (2018)}} \\
& & \textbf{RPS} & \textbf{$\text{Pseudo-R}^2$} & \textbf{RPS} & \textbf{$\text{Pseudo-R}^2$} & \textbf{RPS} & \textbf{$\text{Pseudo-R}^2$} \\
\midrule
Bundesliga & Bivariate Poisson     & \B{0.216} & \B{0.328} & 0.219 & 0.324 & 0.217 & 0.325 \\
 & Diag. Infl. Bivariate Poisson  & 0.245 & 0.298 & 0.242 & 0.302 & 0.238 & 0.306 \\
 & Double Poisson   & 0.235 & 0.307 & 0.236 & 0.309 & 0.227 & 0.317 \\
 & Negative Binomial & 0.225 & 0.316 & 0.229 & 0.314 & 0.220 & 0.322 \\
 & Skellam Model & 0.220 & 0.321 & 0.221 & 0.321 & 0.217 & 0.322 \\
 & Zero Infl. Skellam Model & 0.222 & 0.318 & 0.222 & 0.321 & 0.217 & 0.322 \\
 \midrule
EPL & Bivariate Poisson     & 0.225 & \B{0.370} & 0.228 & 0.367 & 0.228 & 0.366 \\
 & Diag. Infl. Bivariate Poisson  & \B{0.224} & 0.369 & 0.232 & 0.362 & 0.233 & 0.360 \\
 & Double Poisson   & 0.229 & 0.364 & 0.234 & 0.360 & 0.235 & 0.356 \\
 & Negative Binomial & 0.228 & 0.368 & 0.231 & 0.364 & 0.231 & 0.365 \\
 & Skellam Model & 0.230 & 0.360 & 0.234 & 0.358 & 0.232 & 0.362 \\
 & Zero Infl. Skellam Model & 0.228 & 0.361 & 0.236 & 0.358 & 0.235 & 0.358 \\
\midrule
La Liga & Bivariate Poisson     & 0.190 & 0.420 & 0.199 & 0.408 & 0.198 & 0.410 \\
 & Diag. Infl. Bivariate Poisson  & \B{0.189} & \B{0.421} & 0.198 & 0.409 & 0.200 & 0.406 \\
 & Double Poisson   & 0.190 & 0.419 & 0.199 & 0.407 & 0.199 & 0.410 \\
 & Negative Binomial & 0.194 & 0.413 & 0.204 & 0.401 & 0.204 & 0.400 \\
 & Skellam Model & 0.211 & 0.392 & 0.213 & 0.391 & 0.214 & 0.390 \\
 & Zero Infl. Skellam Model & 0.209 & 0.394 & 0.213 & 0.390 & 0.214 & 0.390 \\
\bottomrule
\end{tabular}
\end{adjustbox}
\end{table}

Table~\ref{tab_rps_pseudoR_half_matches} presents the same comparisons for the second half of last season. In the Bundesliga, the weighted dynamic bivariate Poisson model exhibits the lowest RPS $(0.222)$ and the highest $\text{Pseudo-R}^{2}$ of $0.336$. In the EPL, the weighted dynamic bivariate Poisson presents the lowest RPS $(0.205)$ and the highest $\text{Pseudo-R}^{2}$ $(0.380)$. In La Liga, the weighted dynamic version of the bivariate Poisson, Skellam model, and zero-inflated Skellam models achieve an RPS value of $0.200$ and present the largest $\text{Pseudo-R}^{2}$ $(0.375)$.
\begin{table}[htbp!]
\caption{Ranked Probability Score (RPS) and $\text{Pseudo-R}^{2}$ for the proposed weighted dynamic method, \citet{owen_2011} method and \citet{egidi_etal2018} method, evaluated on the second half of the 2024/2025 season for for the Bundesliga, English Premier League (EPL), and La Liga.}
\label{tab_rps_pseudoR_half_matches}
\centering
\begin{adjustbox}{max width=\textwidth}
\begin{tabular}{l
                l
                S[table-format=1.4, table-space-text-pre={\B{0}}] S[table-format=1.4, table-space-text-pre={\B{0}}]
                S[table-format=1.4, table-space-text-pre={\B{0}}] S[table-format=1.4, table-space-text-pre={\B{0}}]
                S[table-format=1.4, table-space-text-pre={\B{0}}] S[table-format=1.4, table-space-text-pre={\B{0}}]}
\toprule
\textbf{League} & \textbf{Model} & 
\multicolumn{2}{c}{\textbf{Weighted Dynamic}} & 
\multicolumn{2}{c}{\textbf{Owen (2011)}} & 
\multicolumn{2}{c}{\textbf{Egidi et al. (2018)}} \\
& & \textbf{RPS} & \textbf{$\text{Pseudo-R}^{2}$} & \textbf{RPS} & \textbf{$\text{Pseudo-R}^{2}$} & \textbf{RPS} & \textbf{$\text{Pseudo-R}^{2}$} \\
\midrule
Bundesliga & Bivariate Poisson     & \B{0.222} & \B{0.336} & 0.224 & 0.334 & 0.223 & 0.335 \\
 & Diag. Infl. Bivariate Poisson  & 0.237 & 0.319 & 0.237 & 0.321 & 0.237 & 0.320 \\
 & Double Poisson   & 0.229 & 0.328 & 0.234 & 0.325 & 0.234 & 0.324 \\
 & Negative Binomial & 0.226 & 0.330 & 0.232 & 0.327 & 0.232 & 0.326 \\
 & Skellam Model & 0.225 & 0.333 & 0.228 & 0.331 & 0.226 & 0.332 \\
 & Zero Infl. Skellam Model & 0.225 & 0.334 & 0.228 & 0.332 & 0.226 & 0.332 \\
 \midrule
EPL & Bivariate Poisson     & \B{0.205} & \B{0.380} & 0.206 & 0.378 & 0.207 & 0.378 \\
 & Diag. Infl. Bivariate Poisson  & 0.212 & 0.371 & 0.210 & 0.371 & 0.209 & 0.375 \\
 & Double Poisson   & 0.208 & 0.377 & 0.207 & 0.377 & 0.206  & 0.378 \\
 & Negative Binomial & 0.208 & 0.376 & 0.206 & 0.377 & 0.207 & 0.376 \\
 & Skellam Model & 0.213 & 0.367 & 0.213 & 0.367 & 0.211 & 0.369 \\
 & Zero Infl. Skellam Model & 0.214 & 0.365 & 0.212 & 0.369 & 0.211 & 0.369 \\
\midrule
La Liga & Bivariate Poisson     & \B{0.200} & \B{0.375} & 0.202 & 0.373 & 0.201 & 0.374 \\
 & Diag. Infl. Bivariate Poisson  & 0.200 & 0.374 & 0.202 & 0.374 & 0.202 & 0.373 \\
 & Double Poisson   & 0.200 & 0.374 & 0.201 & 0.374 & 0.201 & 0.374 \\
 & Negative Binomial & 0.202 & 0.372 & 0.204 & 0.370 & 0.202 & 0.372 \\
 & Skellam Model & \B{0.200} & \B{0.375} & 0.203 & 0.372 & 0.206 & 0.369 \\
& Zero Infl. Skellam Model & \B{0.200} & \B{0.375} & 0.203 & 0.373 & 0.205 & 0.371 \\
\bottomrule
\end{tabular}
\end{adjustbox}
\end{table}



\section{Further analysis on $\phi_{\text{att}}$ and  $\phi_{\text{def}}$ }
\label{app_comm_par}
\begin{figure}[htb!]
    \centering   \includegraphics[width=0.99\textwidth, height=12cm]{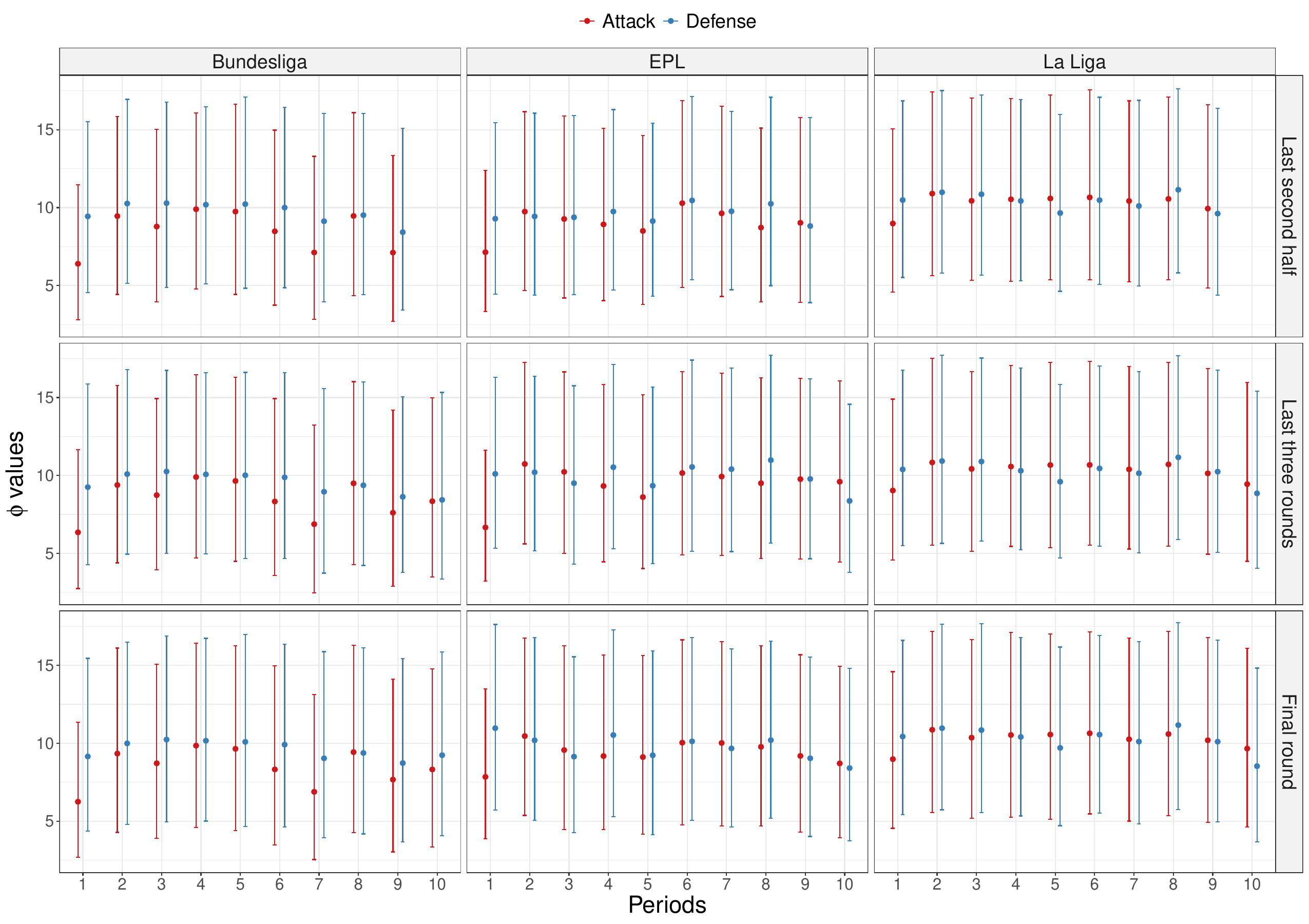}
\caption{Posterior means (points) and $95\%$ credible intervals (error bars) of the commensurate parameters for offensive (red) and defensive (blue) abilities, under the three predictive scenarios for the Bundesliga, English Premier League (EPL), and La Liga.}
\label{comm_par_post}
\end{figure}

Figure \ref{comm_par_post} shows how the posterior commensurate parameters $\boldsymbol{\phi}_{\text{att}}$ and $\boldsymbol{\phi}_{\text{def}}$ for the attacking and defensive abilities of the teams evolve during the evaluated periods, under the three predictive scenarios and for each of the three major European leagues. Based on the result obtained in Section \ref{sec_brier_RPS}, when predicting the final round of the 2024/2025 season, the Bundesliga is modelled with the bivariate Poisson, the EPL with the Skellam model, and La Liga with the diagonal‑inflated bivariate Poisson; when forecasting the last three rounds, the Bundesliga again uses the bivariate Poisson while both the EPL and La Liga use the diagonal‑inflated bivariate Poisson; and when forecasting the remaining half‑season, the best model for the Bundesliga and EPL is the bivariate Poisson, while in La Liga it is the diagonal‑inflated bivariate Poisson.

\section{Computational performance and convergence diagnostics}
\label{app_comp_conv}
\begin{figure}[htb!]
    \centering   \includegraphics[width=0.99\textwidth, height=14cm]{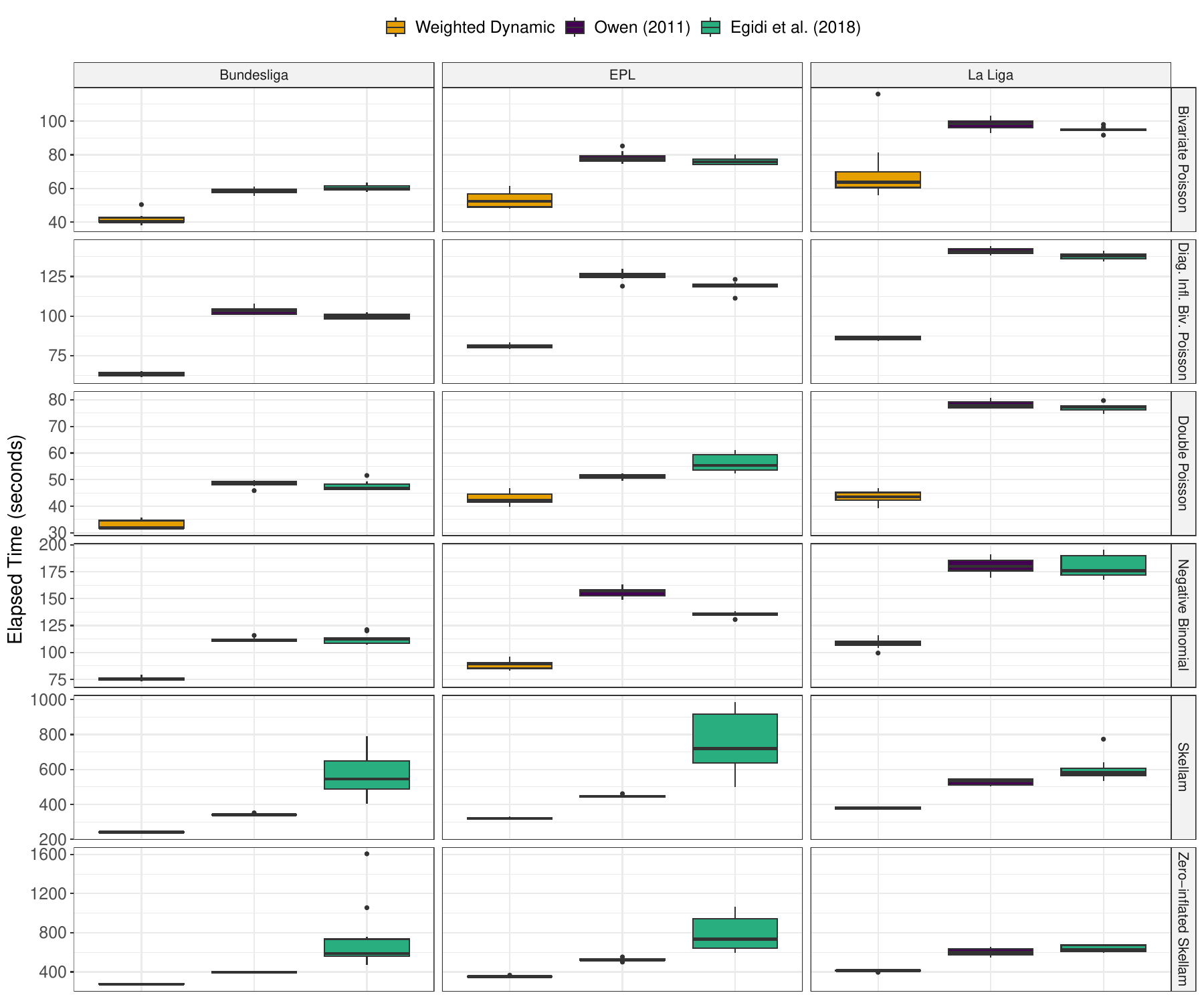}
\caption{Boxplots of elapsed computation times (in seconds).  Results from the proposed weighted dynamic method are shown alongside corresponding values from \citet{owen_2011} and \citet{egidi_etal2018}, all evaluated at the final round of the 2024/2025 season scenario.}
\label{comp_boxplot}
\end{figure}

In terms of computation time, the proposed weighted dynamic approach outperforms the dynamic formulations of \citet{owen_2011} and \citet{egidi_etal2018} in all scenarios. Figure \ref{comp_boxplot} shows the distribution of the elapsed computation times (in seconds) for each method, divided by league and by the six goal-based models considered, evaluated in the final round of the 2024/2025 season scenario. Specifically, for each model–league pair we performed ten independent MCMC fits with four independent and parallel chains, each consisting of 2000 iterations, with the initial 1000 iterations discarded as burn-in.  
Across all leagues and models, the weighted dynamic model consistently requires the least computation time to reach convergence. The median run-time under the weighted dynamic specification is lower than those of \citet{owen_2011} and \citet{egidi_etal2018} methods for every combination of league and model. 
Notably, in the most computationally demanding setting – the zero-inflated Skellam model – our weighted dynamic model for the EPL is $32\%$ faster than Owen's version and $55\%$ faster than \citet{egidi_etal2018} approach. Even for relatively simpler models such as the double Poisson, the weighted dynamic version reduces the runtime by approximately $32\%$ with respect to \citet{owen_2011} and by $31\%$ with respect to \citet{egidi_etal2018} for the Bundesliga. In La Liga the weighted dynamic negative binomial model is $40\%$ faster than \citet{owen_2011} and \citet{egidi_etal2018}.

All three fitting methods achieved satisfactory convergence diagnostics for all the evaluated scenarios. In particular, we verified that the MCMC chains for each model and method yielded Gelman-Rubin statistic $\hat{R}$ \citep{gelman1992inference} very close to 1.00 and large bulk and tail effective sample sizes, indicating stable convergence. Table \ref{tab_convergence} shows a summary of the convergence metrics of the weighted dynamic method for the final
round of the 2024/2025 season scenario. Specifically, the means of the Gelman-Rubin statistic $\hat{R}$, bulk and tail effective sample sizes is shown for the $\boldsymbol{\beta}^{\text{att}}, \boldsymbol{\beta}^{\text{def}}, \boldsymbol{\phi}_{\text{att}}$ and $\boldsymbol{\phi}_{\text{def}}$ parameters. This suggests that the substantial computational efficiency of the weighted dynamic approach does not come at the expense of sampler stability or accuracy. Thus, the weighted dynamic model not only provides flexibility for dynamic team-specific parameters, but does so with a lower computational cost.
\begin{table}[htbp!]
\caption{%
  MCMC convergence diagnostics: $\hat{R}$, bulk and tail effective sample sizes (ESS) mean of the
   $\boldsymbol{\beta}^{\text{att}}, \boldsymbol{\beta}^{\text{def}}, \boldsymbol{\phi}_{\text{att}}$ and $\boldsymbol{\phi}_{\text{def}}$ parameters for the proposed weighted dynamic method, all evaluated at the final round of the 2024/2025 season scenario. 
}
\label{tab_convergence}
\centering
\begin{adjustbox}{max width=\textwidth}
\begin{tabular}{
  ll
  S[table-format=1.2] S[table-format=4.0] S[table-format=4.0]
  S[table-format=1.2] S[table-format=4.0] S[table-format=4.0]
  S[table-format=1.2] S[table-format=4.0] S[table-format=4.0]
  S[table-format=1.2] S[table-format=4.0] S[table-format=4.0]
}
\toprule
\multicolumn{2}{c}{} 
  & \multicolumn{3}{c}{\textbf{\(\beta^{\text{att}}\)}} 
  & \multicolumn{3}{c}{\textbf{\(\beta^{\text{def}}\)}} 
  & \multicolumn{3}{c}{\textbf{\(\phi_{\text{att}}\)}} 
  & \multicolumn{3}{c}{\textbf{\(\phi_{\text{def}}\)}} \\
\cmidrule(lr){3-5} \cmidrule(lr){6-8} \cmidrule(lr){9-11} \cmidrule(lr){12-14}
\textbf{League} & \textbf{Model} 
  & \textbf{\(\bar{\hat R}\)} & \textbf{$\overline{\text{Bulk ESS}}$} & \textbf{$\overline{\text{Tail ESS}}$}
  & \textbf{\(\bar{\hat R}\)} & \textbf{$\overline{\text{Bulk ESS}}$} & \textbf{$\overline{\text{Tail ESS}}$}
  & \textbf{\(\bar{\hat R}\)} & \textbf{$\overline{\text{Bulk ESS}}$} & \textbf{$\overline{\text{Tail ESS}}$}
  & \textbf{\(\bar{\hat R}\)} & \textbf{$\overline{\text{Bulk ESS}}$} & \textbf{$\overline{\text{Tail ESS}}$} \\
\midrule
Bundesliga & Bivariate Poisson     
  & 1.00 & 4358 & 2889   
  & 1.00 & 4482 & 2960   
  & 1.00 & 2980 & 2742       
  & 1.00 & 3202 & 2711       \\
& Diag.\ Infl.\ Biv.\ Poisson  
  & 1.00 & 4728       & 3020   
  & 1.00 & 4997       & 3061   
  & 1.00 & 3882       & 2940       
  & 1.00 & 4073       & 3157       \\
             & Double Poisson   
  & 1.00 & 4917       & 3008   
  & 1.00 & 5085       & 3012   
  & 1.00 & 4186      & 3032       
  & 1.00 & 4089       & 2839       \\
             & Negative Binomial  
  & 1.00 & 4036       & 3225   
  & 1.00 & 4993       & 3101   
  & 1.00 & 3658       & 2960       
  & 1.00 & 3878       & 2900       \\
             & Skellam Model  
  & 1.00 & 4772       & 3550   
  & 1.00 & 4802       & 3502   
  & 1.00 & 3370       & 3411       
  & 1.00 & 3431       & 3187       \\
             & Zero Infl. Skellam Model   
  & 1.00 & 4915       & 3554   
  & 1.00 & 4962       & 3567   
  & 1.00 & 3384       & 3338       
  & 1.00 & 3424       & 3287     \\
  \midrule
EPL & Bivariate Poisson     
  & 1.00 & 4473 & 2903   
  & 1.00 & 4580      & 2980   
  & 1.00 & 2948      & 2671       
  & 1.00 & 3240       & 2858       \\
           & Diag.\ Infl.\ Biv.\ Poisson  
  & 1.00 & 4637       & 3016   
  & 1.00 & 4805       & 2983   
  & 1.00 & 3462       & 2981       
  & 1.00 & 3675       & 2832       \\
             & Double Poisson  
  & 1.00 & 4938       & 2951   
  & 1.00 & 5197       & 2960   
  & 1.00 & 3888       & 3114       
  & 1.00 & 3752       & 2931       \\
             & Negative Binomial  
  & 1.00 & 3560       & 3024   
  & 1.00 & 4824       & 3004   
  & 1.00 & 3584       & 3100       
  & 1.00 & 3552       & 3006       
         \\
             & Skellam Model  
  & 1.00 & 4253       & 3299   
  & 1.00 & 4344       & 3312   
  & 1.00 & 3415      & 3122       
  & 1.00 & 3189      & 2978       \\
             & Zero Infl. Skellam Model  
  & 1.00 & 4478       & 3370  
  & 1.00 & 4621       & 3380   
  & 1.00 & 3444       & 3235       
  & 1.00 & 3244       & 3074      \\
  \midrule
La Liga & Bivariate Poisson     
  & 1.00 & 3898 & 2818   
  & 1.00 & 4100       & 2886   
  & 1.00 & 2993       & 2723       
  & 1.00 & 2933       & 2627       \\
           & Diag.\ Infl.\ Biv.\ Poisson  
  & 1.00 & 4289       & 2980   
  & 1.00 & 4583       & 3027   
  & 1.00 & 3805       & 2891       
  & 1.00 & 3592       & 2942       \\
             & Double Poisson  
  & 1.00 & 4655       & 2942   
  & 1.00 & 4820       & 2971   
  & 1.00 & 4099       & 3121       
  & 1.00 & 4100      & 3035      \\
             & Negative Binomial  
  & 1.00 & 3059       & 2917   
  & 1.00 & 4523       & 2911   
  & 1.00 & 3844       & 2878       
  & 1.00 & 3615       & 2887       \\
             & Skellam Model  
  & 1.00 & 3683       & 3168   
  & 1.00 & 3752       & 3187   
  & 1.00 & 3187       & 2793       
  & 1.00 & 3170       & 2978       \\
             & Zero Infl. Skellam Model  
  & 1.00 & 4056       & 3287   
  & 1.00 & 4062       & 3263   
  & 1.00 & 3382       & 3012       
  & 1.00 & 3535       & 3179   \\
  \bottomrule

\end{tabular}
\end{adjustbox}
\end{table}

\begin{figure}[htbp!]
    \centering   \includegraphics[width=0.99\textwidth, height=14cm]{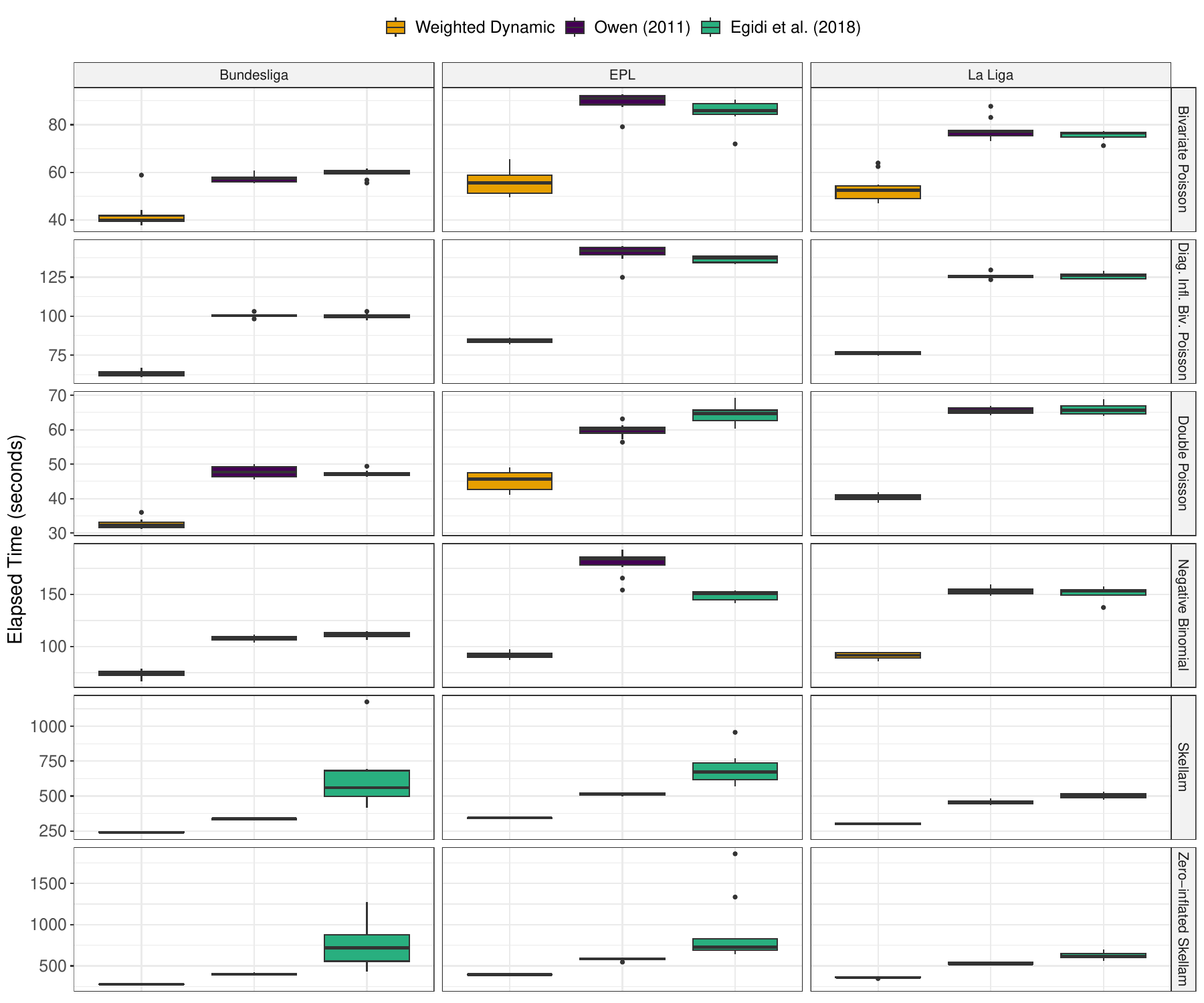}
\caption{Boxplots of elapsed computation times (in seconds).  Results from the proposed weighted dynamic method are shown alongside corresponding values from \citet{owen_2011} and \citet{egidi_etal2018}, all evaluated at the last three matches of the 2024/2025 season scenario.}
\label{comp_boxplot_half_30}
\end{figure}

\begin{figure}[htb!]
    \centering   \includegraphics[width=0.99\textwidth, height=14cm]{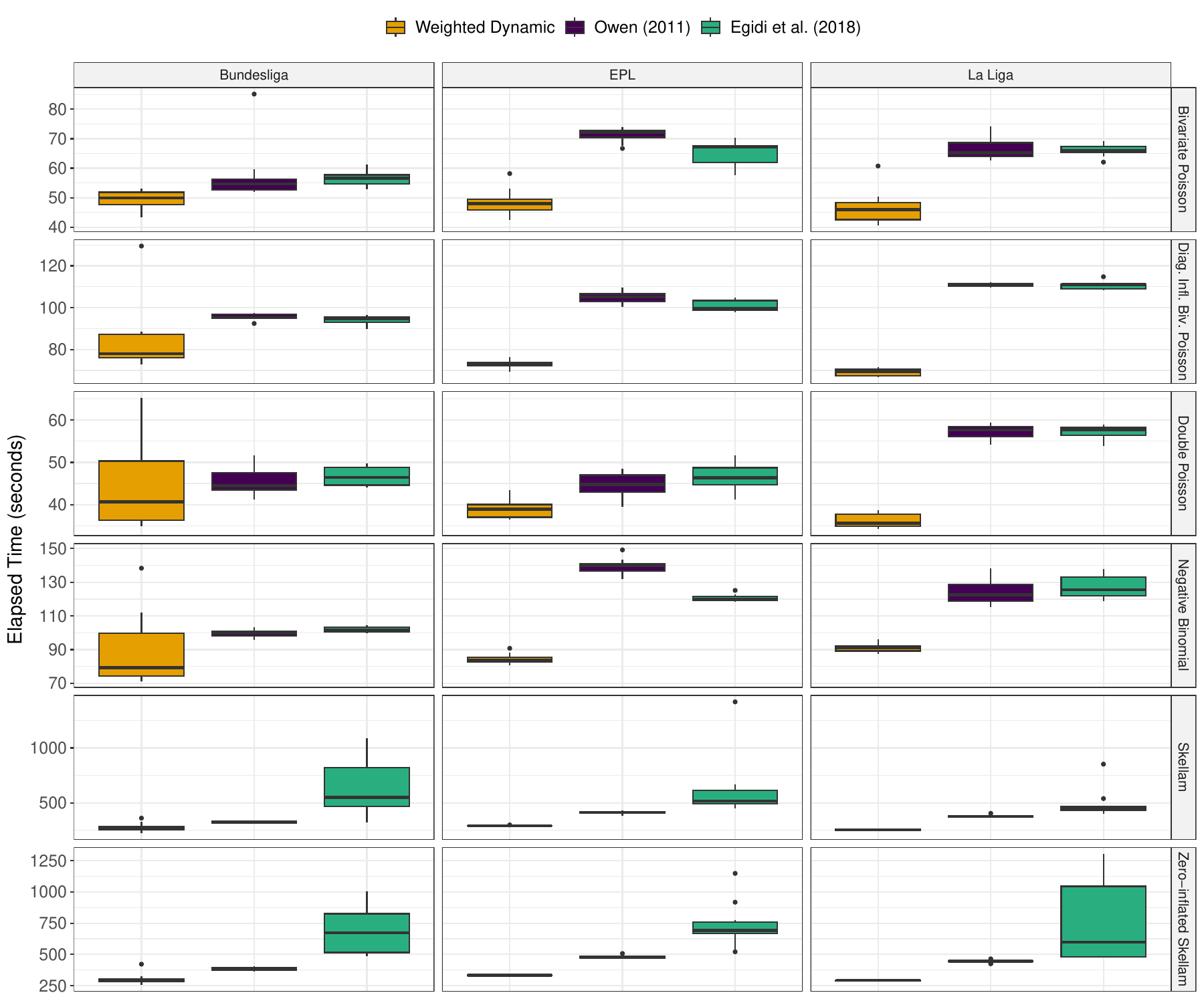}
\caption{Boxplots of elapsed computation times (in seconds).  Results from the proposed weighted dynamic method are shown alongside corresponding values from \citet{owen_2011} and \citet{egidi_etal2018}, all evaluated at the second half of the 2024/2025 season scenario.}
\label{comp_boxplot_half}
\end{figure}

Similar analyses on computational time and convergence for the other two predictive scenarios (last three rounds and the second half of the 2024/2025 season) are presented in Figures \ref{comp_boxplot_half_30} and \ref{comp_boxplot_half}, as well as Tables \ref{tab_convergence_half_30} and \ref{tab_convergence_half}.

\begin{table}[htbp!]
\caption{%
  MCMC convergence diagnostics: $\hat{R}$, bulk and tail effective sample sizes (ESS) mean of the
   $\boldsymbol{\beta}^{\text{att}}, \boldsymbol{\beta}^{\text{def}}, \boldsymbol{\phi}_{\text{att}}$ and $\boldsymbol{\phi}_{\text{def}}$ parameters for the proposed weighted dynamic method, all evaluated at the last three round of the 2024/2025 season scenario. 
}
\label{tab_convergence_half_30}
\centering
\begin{adjustbox}{max width=\textwidth}
\begin{tabular}{
  ll
  S[table-format=1.2] S[table-format=4.0] S[table-format=4.0]
  S[table-format=1.2] S[table-format=4.0] S[table-format=4.0]
  S[table-format=1.2] S[table-format=4.0] S[table-format=4.0]
  S[table-format=1.2] S[table-format=4.0] S[table-format=4.0]
}
\toprule
\multicolumn{2}{c}{} 
  & \multicolumn{3}{c}{\textbf{\(\beta^{\text{att}}\)}} 
  & \multicolumn{3}{c}{\textbf{\(\beta^{\text{def}}\)}} 
  & \multicolumn{3}{c}{\textbf{\(\phi_{\text{att}}\)}} 
  & \multicolumn{3}{c}{\textbf{\(\phi_{\text{def}}\)}} \\
\cmidrule(lr){3-5} \cmidrule(lr){6-8} \cmidrule(lr){9-11} \cmidrule(lr){12-14}
\textbf{League} & \textbf{Model} 
  & \textbf{\(\bar{\hat R}\)} & \textbf{$\overline{\text{Bulk ESS}}$} & \textbf{$\overline{\text{Tail ESS}}$}
  & \textbf{\(\bar{\hat R}\)} & \textbf{$\overline{\text{Bulk ESS}}$} & \textbf{$\overline{\text{Tail ESS}}$}
  & \textbf{\(\bar{\hat R}\)} & \textbf{$\overline{\text{Bulk ESS}}$} & \textbf{$\overline{\text{Tail ESS}}$}
  & \textbf{\(\bar{\hat R}\)} & \textbf{$\overline{\text{Bulk ESS}}$} & \textbf{$\overline{\text{Tail ESS}}$} \\
\midrule
Bundesliga & Bivariate Poisson     
  & 1.00 & 4532       & 2975   
  & 1.00 & 4908       & 3048   
  & 1.00 & 3206       & 2813       
  & 1.00 & 3436       & 3008      \\
& Diag.\ Infl.\ Biv.\ Poisson  
  & 1.00 & 4722       & 3046   
  & 1.00 & 4796       & 3014   
  & 1.00 & 3604       & 2972       
  & 1.00 & 3830       & 2987       \\
             & Double Poisson   
  & 1.00 & 5013       & 3005   
  & 1.00 & 5258       & 3040   
  & 1.00 & 4058      & 2998       
  & 1.00 & 4248       & 3021       \\
             & Negative Binomial  
  & 1.00 & 3709       & 3111   
  & 1.00 & 5273       & 3064   
  & 1.00 & 3870       & 2979       
  & 1.00 & 3984       & 3023       \\
             & Skellam Model  
  & 1.00 & 4724       & 3469   
  & 1.00 & 4841       & 3492   
  & 1.00 & 3308       & 3214       
  & 1.00 & 3254       & 2999       \\
             & Zero Infl. Skellam Model   
  & 1.00 & 4788       & 3552   
  & 1.00 & 4862       & 3557   
  & 1.00 & 3364       & 3254       
  & 1.00 & 3481       & 3337     \\
  \midrule
EPL & Bivariate Poisson     
  & 1.00 & 4302 & 2885   
  & 1.00 & 4413      & 2920   
  & 1.00 & 3079      & 2829       
  & 1.00 & 3002       & 2832       \\
           & Diag.\ Infl.\ Biv.\ Poisson  
  & 1.00 & 4635       & 2990   
  & 1.00 & 4785       & 2989   
  & 1.00 & 3434       & 3057       
  & 1.00 & 3529       & 2957       \\
             & Double Poisson  
  & 1.00 & 4868       & 2978   
  & 1.00 & 5008       & 2964   
  & 1.00 & 3647       & 2841       
  & 1.00 & 3939       & 2914       \\
             & Negative Binomial  
  & 1.00 & 3536       & 3022   
  & 1.00 & 4705       & 2963   
  & 1.00 & 3540       & 2868       
  & 1.00 & 3427       & 2675       \\
             & Skellam Model  
  & 1.00 & 4643       & 3447   
  & 1.00 & 4781       & 3457   
  & 1.00 & 3456      & 3226       
  & 1.00 & 3390       & 3116       \\
             & Zero Infl. Skellam Model  
  & 1.00 & 4628       & 3409  
  & 1.00 & 4772       & 3440   
  & 1.00 & 3241       & 3206       
  & 1.00 & 3338       & 3084      \\
  \midrule
La Liga & Bivariate Poisson     
  & 1.00 & 4129 & 2888   
  & 1.00 & 4026       & 2884   
  & 1.00 & 2894       & 2843       
  & 1.00 & 3037       & 2710       \\
           & Diag.\ Infl.\ Biv.\ Poisson  
  & 1.00 & 4451       & 3016   
  & 1.00 & 4584       & 3000   
  & 1.00 & 3753       & 2999       
  & 1.00 & 3651       & 2940       \\
             & Double Poisson  
  & 1.00 & 4839       & 2969   
  & 1.00 & 4863       & 2938   
  & 1.00 & 4056       & 2887       
  & 1.00 & 4029      & 2939      \\
             & Negative Binomial  
  & 1.00 & 2873       & 2821   
  & 1.00 & 4283       & 2892   
  & 1.00 & 3862       & 2964       
  & 1.00 & 3562       & 2886       \\
             & Skellam Model  
  & 1.00 & 4210       & 3303   
  & 1.00 & 4288       & 3292   
  & 1.00 & 3592       & 3101       
  & 1.00 & 3450       & 2919       \\
             & Zero Infl. Skellam Model  
  & 1.00 & 4259       & 3307   
  & 1.00 & 4342       & 3317   
  & 1.00 & 3221       & 3005       
  & 1.00 & 3289       & 2957   \\
  \bottomrule

\end{tabular}
\end{adjustbox}
\end{table}

\begin{table}[htbp!]
\caption{%
  MCMC convergence diagnostics: $\hat{R}$, bulk and tail effective sample sizes (ESS) mean of the
   $\boldsymbol{\beta}^{\text{att}}, \boldsymbol{\beta}^{\text{def}}, \boldsymbol{\phi}_{\text{att}}$ and $\boldsymbol{\phi}_{\text{def}}$ parameters for the proposed weighted dynamic method, all evaluated at the second half of the 2024/2025 season scenario. 
}
\label{tab_convergence_half}
\centering
\begin{adjustbox}{max width=\textwidth}
\begin{tabular}{
  ll
  S[table-format=1.2] S[table-format=4.0] S[table-format=4.0]
  S[table-format=1.2] S[table-format=4.0] S[table-format=4.0]
  S[table-format=1.2] S[table-format=4.0] S[table-format=4.0]
  S[table-format=1.2] S[table-format=4.0] S[table-format=4.0]
}
\toprule
\multicolumn{2}{c}{} 
  & \multicolumn{3}{c}{\textbf{\(\beta^{\text{att}}\)}} 
  & \multicolumn{3}{c}{\textbf{\(\beta^{\text{def}}\)}} 
  & \multicolumn{3}{c}{\textbf{\(\phi_{\text{att}}\)}} 
  & \multicolumn{3}{c}{\textbf{\(\phi_{\text{def}}\)}} \\
\cmidrule(lr){3-5} \cmidrule(lr){6-8} \cmidrule(lr){9-11} \cmidrule(lr){12-14}
\textbf{League} & \textbf{Model} 
  & \textbf{\(\bar{\hat R}\)} & \textbf{$\overline{\text{Bulk ESS}}$} & \textbf{$\overline{\text{Tail ESS}}$}
  & \textbf{\(\bar{\hat R}\)} & \textbf{$\overline{\text{Bulk ESS}}$} & \textbf{$\overline{\text{Tail ESS}}$}
  & \textbf{\(\bar{\hat R}\)} & \textbf{$\overline{\text{Bulk ESS}}$} & \textbf{$\overline{\text{Tail ESS}}$}
  & \textbf{\(\bar{\hat R}\)} & \textbf{$\overline{\text{Bulk ESS}}$} & \textbf{$\overline{\text{Tail ESS}}$} \\
\midrule
Bundesliga & Bivariate Poisson     
  & 1.00 & 4980 & 3057   
  & 1.00 & 5358 & 3135   
  & 1.00 & 3225 & 2945       
  & 1.00 & 3584 & 2998       \\
& Diag.\ Infl.\ Biv.\ Poisson  
  & 1.00 & 4784       & 3157   
  & 1.00 & 4949       & 3128   
  & 1.00 & 3602       & 2981       
  & 1.00 & 3552       & 2974       \\
             & Double Poisson   
  & 1.00 & 5298       & 3158   
  & 1.00 & 5502       & 3181   
  & 1.00 & 4151      & 3136       
  & 1.00 & 4194       & 3099       \\
             & Negative Binomial  
  & 1.00 & 4517       & 3365   
  & 1.00 & 5063       & 3218   
  & 1.00 & 3877       & 3129       
  & 1.00 & 3789       & 3151       \\
             & Skellam Model  
  & 1.00 & 4736       & 3501   
  & 1.00 & 4811       & 3536   
  & 1.00 & 3495       & 3385       
  & 1.00 & 3410       & 3324       \\
             & Zero Infl. Skellam Model   
  & 1.00 & 4830       & 3592   
  & 1.00 & 4832       & 3562   
  & 1.00 & 3432       & 3372       
  & 1.00 & 3383       & 3340     \\
  \midrule
EPL & Bivariate Poisson     
  & 1.00 & 4894 & 2994   
  & 1.00 & 5053      & 2971   
  & 1.00 & 3152      & 2855       
  & 1.00 & 3205       & 2841       \\
           & Diag.\ Infl.\ Biv.\ Poisson  
  & 1.00 & 5361       & 3112   
  & 1.00 & 5581       & 3096   
  & 1.00 & 3831       & 3102       
  & 1.00 & 3856       & 3145       \\
             & Double Poisson  
  & 1.00 & 5288       & 3017   
  & 1.00 & 5561       & 3023   
  & 1.00 & 3968       & 3003       
  & 1.00 & 3855       & 3083       \\
             & Negative Binomial  
  & 1.00 & 4330       & 3294   
  & 1.00 & 5161       & 3199   
  & 1.00 & 3605       & 2939       
  & 1.00 & 3642       & 3191       \\
             & Skellam Model  
  & 1.00 & 4707       & 3432   
  & 1.00 & 4828       & 3475   
  & 1.00 & 3403      & 3192       
  & 1.00 & 3430       & 3147       \\
             & Zero Infl. Skellam Model  
  & 1.00 & 4610       & 3374  
  & 1.00 & 4722       & 3407   
  & 1.00 & 3358       & 3199       
  & 1.00 & 3299       & 3062      \\
  \midrule
La Liga & Bivariate Poisson     
  & 1.00 & 3898 & 2818   
  & 1.00 & 4100       & 2886   
  & 1.00 & 2993       & 2723       
  & 1.00 & 2993       & 2627       \\
           & Diag.\ Infl.\ Biv.\ Poisson  
  & 1.00 & 4289       & 2980   
  & 1.00 & 4583       & 3027   
  & 1.00 & 3805       & 2891       
  & 1.00 & 3592       & 2947       \\
             & Double Poisson  
  & 1.00 & 4655       & 2942   
  & 1.00 & 4820       & 2971   
  & 1.00 & 3491       & 3221       
  & 1.00 & 3509      & 3045      \\
             & Negative Binomial  
  & 1.00 & 3059       & 2917   
  & 1.00 & 4523       & 2911   
  & 1.00 & 3844       & 2878       
  & 1.00 & 3615       & 2887       \\
             & Skellam Model  
  & 1.00 & 3683       & 3168   
  & 1.00 & 3752       & 3187   
  & 1.00 & 3187       & 2793       
  & 1.00 & 3170       & 2978       \\
             & Zero Infl. Skellam Model  
  & 1.00 & 4056       & 3287   
  & 1.00 & 4062       & 3263   
  & 1.00 & 3382       & 3012       
  & 1.00 & 3535       & 3179   \\
  \bottomrule

\end{tabular}
\end{adjustbox}
\end{table}

\end{appendices}



\FloatBarrier

\bibliographystyle{apalike}
\bibliography{main}

\end{document}